\documentclass[]{raa}            
\usepackage{graphicx,times}
\usepackage{natbib}
\usepackage{amssymb,amsmath}

\providecommand{\bm}[1]{\mbox{\boldmath$#1$\unboldmath}}

\begin{document}

  \title{Long-term trends in the solar activity.  Variations of solar indices in last 40 years}
 \volnopage{ {\bf 2018} Vol.\ {\bf 00} No. {\bf }, 000--000}
   \setcounter{page}{1}

   \author{E. A. Bruevich
   \inst{},
   \and V. V. Bruevich
   \inst{}
   }

 \institute{Sternberg Astronomical Institute, Moscow State
 University, Universitetsky pr., 13, Moscow 119992, Russia;
            {\it red-field@yandex.ru, brouev@sai.msu.ru,}\\
   \vs \no
   {\small Received [year] [month] [day]; accepted [year] [month] [day] }
}

\abstract{The analysis of long-term variations of several solar activity
indices (AI) since in last 40 years has been performed.
We studied the AI which characterize  the fluxes from  
different areas in solar atmosphere. Our consideration of 
mutual correlations between the solar indices is based on the study
of relationships between them in the period from 1950 to 1990.
This period of time, covering activity cycles 19 -- 22, is
characterized by relatively stable relations between the indices. We
have studied the normalized variations of these indices in the 
recent time in relation to their values which have been calculated
with help of the radiation of the sun in the radio range at a wavelength of 10.7 cm ($F_{10.7}$) in 1950 -- 1990. 
The analysis of time series of variations of the normalized AI (AIFF) in solar cycles
23 -- 24 shows an existence of different trends for different indices in their long-term variations. We assume that variations of normalized SSN, $F_{530.3}$ and Flare Index, which have shown a sharp decrease in last 40 years is possibly associated with a decrease in the intensity of large-scale magnetic fields in the photosphere (SSN) and in the corona (the coronal index and the Flare Index).}

\authorrunning{ E. A. Bruevich, and V. V. Bruevich}            
   \titlerunning{Long-term trends in the solar activity }  
   \maketitle
\keywords{method: data analysis -- Sun: activity indices}

\section{Introduction}
{\label{S:intro}}

A magnetic field of the Sun is the main source of a variability of solar radiation. 
Dominant periodic variations known as Schwabe (11-yr) and Hale (22-yr) cycles show a strong connection with magnetic field's structuring and evolution. A link
between a solar magnetic field  evolution and variation of the AI which are associated with photosphere and corona can be explained by the concept of magnetic helicity. A magnetic helicity is one of the so-called invariants of motion in magnetic hydrodynamics (that is, the physical quantity, the value of which in some physical process does not change over time)
This idea has been used to explain some phenomena which are related on solar activity (Pevtsov et al. 2011).

Variations in solar AI are closely interconnected so as variations of magnetic field are the sources of these changes. Stars of late-type spectral classes are also characterized by a relatively close correlation between the radiation fluxes out coming from star's photosphere, chromosphere and corona (Bruevich \& Alekseev 2007).

In Hathaway (2015); Pevtsov et al. (2014) it was shown that 24th activity cycle is much weaker than cycles 21, 22 and 23.
We can see that according to observations of the International Sunspot Number (SSN)
in cycles
23, 24 and possibly during next cycle 25 the Sun are characterized by low
activity, see Figure 1. 

We see that the current minimum of activity is similar to minimum of Dalton. This minimum was characterized by low solar activity. It corresponds to solar cycles 4 -- 7 (from 1790 to 1830).
  Some scientists also suggested
that soon solar activity according to SSN values in cycles 25
-- 26 will be also very low (SSN will be about 50 -- 70 in the cycle's maximum) 
like low solar activity in Dalton or Maunder minimums (Casey 2014). It is also supposed that this
current minimum is a result of a superimpose on the 11-yr minimum the
minima of cycles with 50-yr and 100-yr periods. (Usoskin \& Mursula 2003) has  also suggested that a long-term decreasing of solar activity is an influence of the current minimum of the century Gleissberg 100-yr cycle.

For sun-like stars of HK project the chromosphere activity cycles were observed at 
 the Mount Wilson observatory during last 50 years.
The analysis of stellar data sets showed that stars with clearly defined
solar-type cycles  have chromospheric fluxes which are corresponding to
the Maunder-like minimum conditions during about 25\% of the time
(Baliunas et al. 1995).
 
We are also nine years into cycle 24. Solar cycle 24 is almost completely passed and the Sun is approaching to its next minimum of activity. The current observed values of maximum amplitudes of solar indices variations show that this cycle is the smallest since cycle 14 (Pesnell 2012). 
In cycle 24 we can see the  "double-peaked" solar maximum. The second maximum was in April 2014 (smoothed values of SSN reached 116.4).  This second peak of SSN surpassed the first peak (smoothed values of SSN reached 98.3 in March 2012). Cycle 24 is the unusual cycle in which the second maximum (according to SSN observations) was larger than SSN in the first maximum.
For SSN it was shown that there is  opposite long-scale
trends during solar cycles 22 -- 24. In (Nagovitsyn et al. 2012) it was analysed last SSN observations from Penn \& Livingston (2006); Pevtsov et al. (2011).
 
Nagovitsyn et al. (2012) has also showed that number of spots with a large areas decreases, and a number of spots with a small areas becomes larger. This fact can be explained by a gradual decrease in magnetic field strength in sunspots. This assumption agrees well with the observation data, see  Penn \& Livingston (2006).

A cycle's evolution can be studied with the examination of variation of the empirical function which connects the SSN and $F_{10.7}$ in cycles 23 -- 24. 

A ratio of the observed sunspot number $SSN^{obs}$ to that
predicted $SSN^{synt}$ using the previous established relationship
to $F_{10.7}$ flux for the period from 1950 to 1990 when these
relationships between solar indices  were stable was called the sunspot
formation fraction -- SFF, which was introduced by Livingston et al. (2012):

\begin{equation} 
SFF = SSN^{obs}/ SSN^{synt}
\end{equation}

For different solar indices in this paper we determine the AIFF -- the activity indices (AI) formation fraction, similar to the SFF parameter, corresponding to
a ratio of the AI observed to the AI  predicted. For the AI  predicted calculations we use
a previous  established relationship to $F_{10.7}$ flux from
the Table 1 below.  We determine AIFF as:

\begin{equation}
 AIFF = F_{ind}^{obs}/F_{ind}^{synt}
\end{equation}

The close relationships between the solar indices observed
from 1950 to 1990 could be considered to be well established. 

The significant deviation from these relationships in cycles 23 -- 24 from previous relationships in 1950 -- 1990 can be used in study of the long-term trends in relationships
between the indices (Livingston et al. 2012).

According to Livingston et al. (2012) the average magnetic field in sunspots in 1990 -- 2008 was decreased by 25\%. Later Watson et al. (2014) examines some of the properties of sunspot umbrae over the 1997 -- 2014 with three different instruments on the ground and in space: MDI, HMI and BABO. It was shown that annual average magnetic field in the darkest part of sunspot umbrae was decreased continuously approximately from 2800 Gauss in 1998 to 2100 Gauss in 2008, and then have  stabilized and remained equal to 2200 Gauss until 2014.

Since magnetic fields in
sunspots decreased so the sunspots are getting not so cold and then their
temperature contrast with a surrounding photosphere (which  is
usually equal to 1500K) getting smaller, making some spots less
visible. So the TSI might  become a little higher, see
 (Svalgaard 2013; Bruevich et al. 2014).

The minimum of solar activity between the cycles 23 and 24
was very long, because of the decline phase of
cycle 23 was unusually long. The solar activity index which was introduced at San Fernando Observatory is based on  photometric images obtained on this observatory,  it represents a  combination of areas of sunspots and faculae.
 Areas of sunspots and faculae are easy to observe, these indices well describe the current state of solar activity.
In Chapman et al. (2014) it was shown that areas of sunspots in cycles 22, 23 and 24 were changed as 1.0, 0.74 and 0.37.

In Figure 1 we see the yearly data of International Sunspot Number for 1700 -- 2017, note that the 1700 -- 1850 data are 
results of indirect assessments. In Figure 1 also we can see  three
well-defined cycles of activity: the main cycle of activity (this
cycle is approximately equal to a 10 -- 11 yrs) and the 40 -- 50 yr cyclicity
with the 100 -- 120 yr (ancient) cyclicity (according to solid line connecting the maxima of the 11-yr cycles). 

 We can also see that the trend shown in Figure 1 (using a thick solid line) 
indicates a progressive decrease of SSN during 50 years gone.
The study of near-surface structural changes in recent cycles show
that there exist the possibility of long-term variation of solar dynamo parameters (Howe et al. 2017).

In Table 1 we see the consolidate information of solar activity indexes and web sites where observational data are available.

\begin{figure}    
   \centerline{\includegraphics[width=100mm]{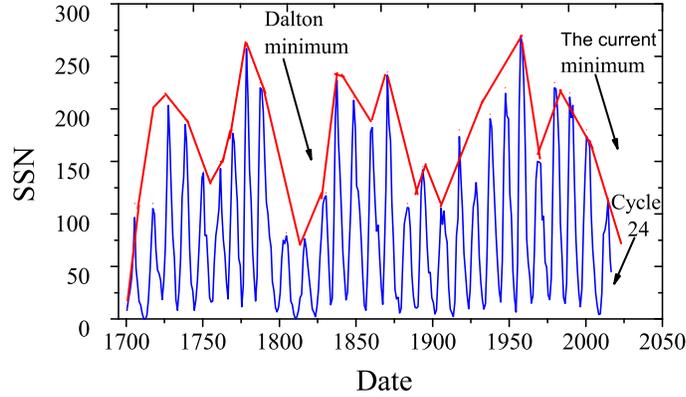}}
\caption{International Sunspot Number -- yearly SSN. Observations for 1700 -- 2017.}
\end{figure}

In  Janardhan et al. (2010) the NSO Kitt Peak synoptic magnetograms were analyzed for cycles 21 -- 23. It was shown that the absolute values of the polar fields has decreased during cycle 23 compared to cycles 21 and 22.

The aim of our paper is an examining of long-term variations of AI which are important in study of solar-terrestrial connections in cycles 21 -- 24. Main trends in long-term changes (on the time scale of 30 -- 50 years) differ for the studied solar indices: this fact shows to various  mechanisms of formation and temporal evolution of solar indices.

\begin{figure}    
   \centerline{\hspace*{0.015\textwidth}
               \includegraphics[width=0.55\textwidth,clip=]{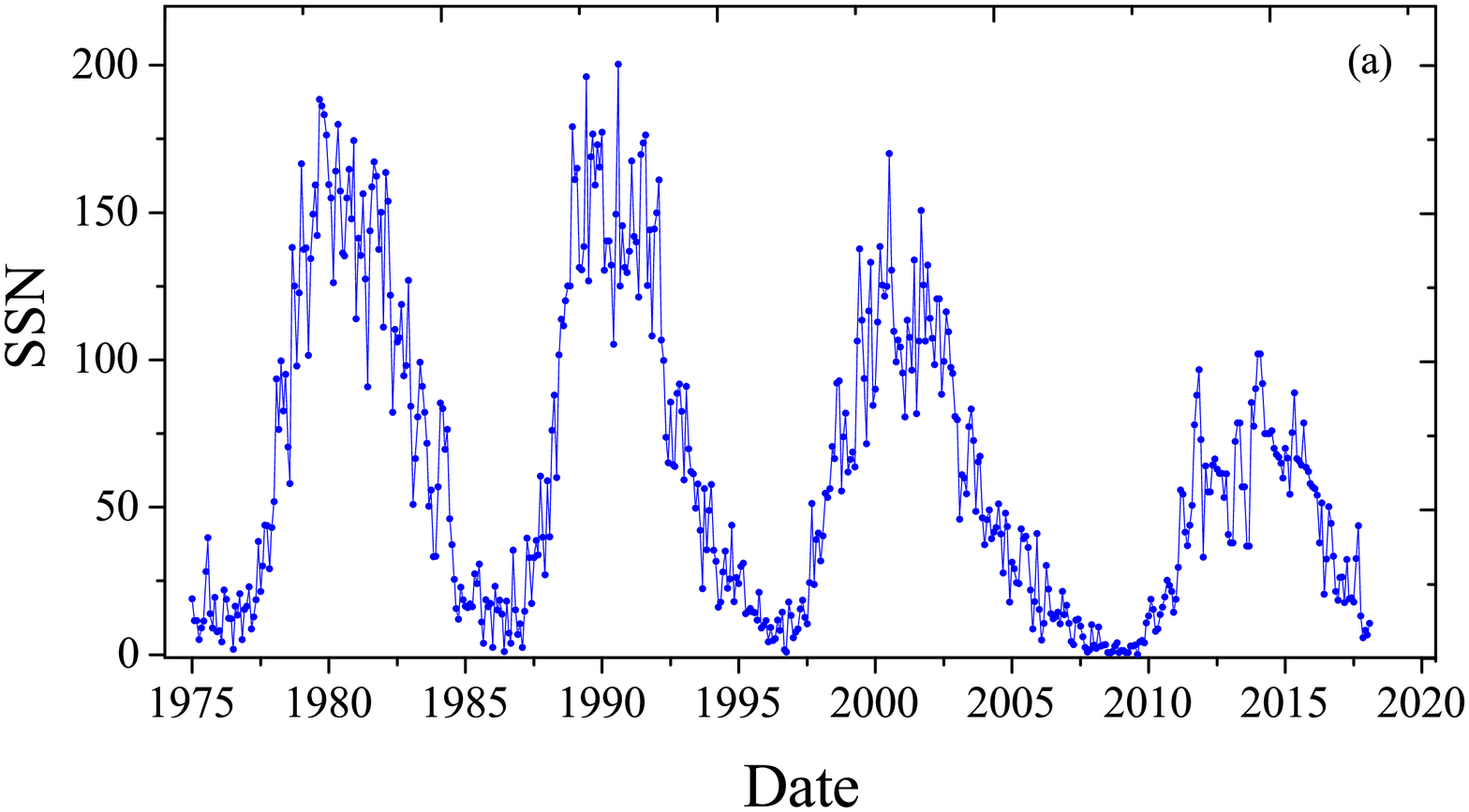}
               \hspace*{-0.015\textwidth}
               \includegraphics[width=0.52\textwidth,clip=]{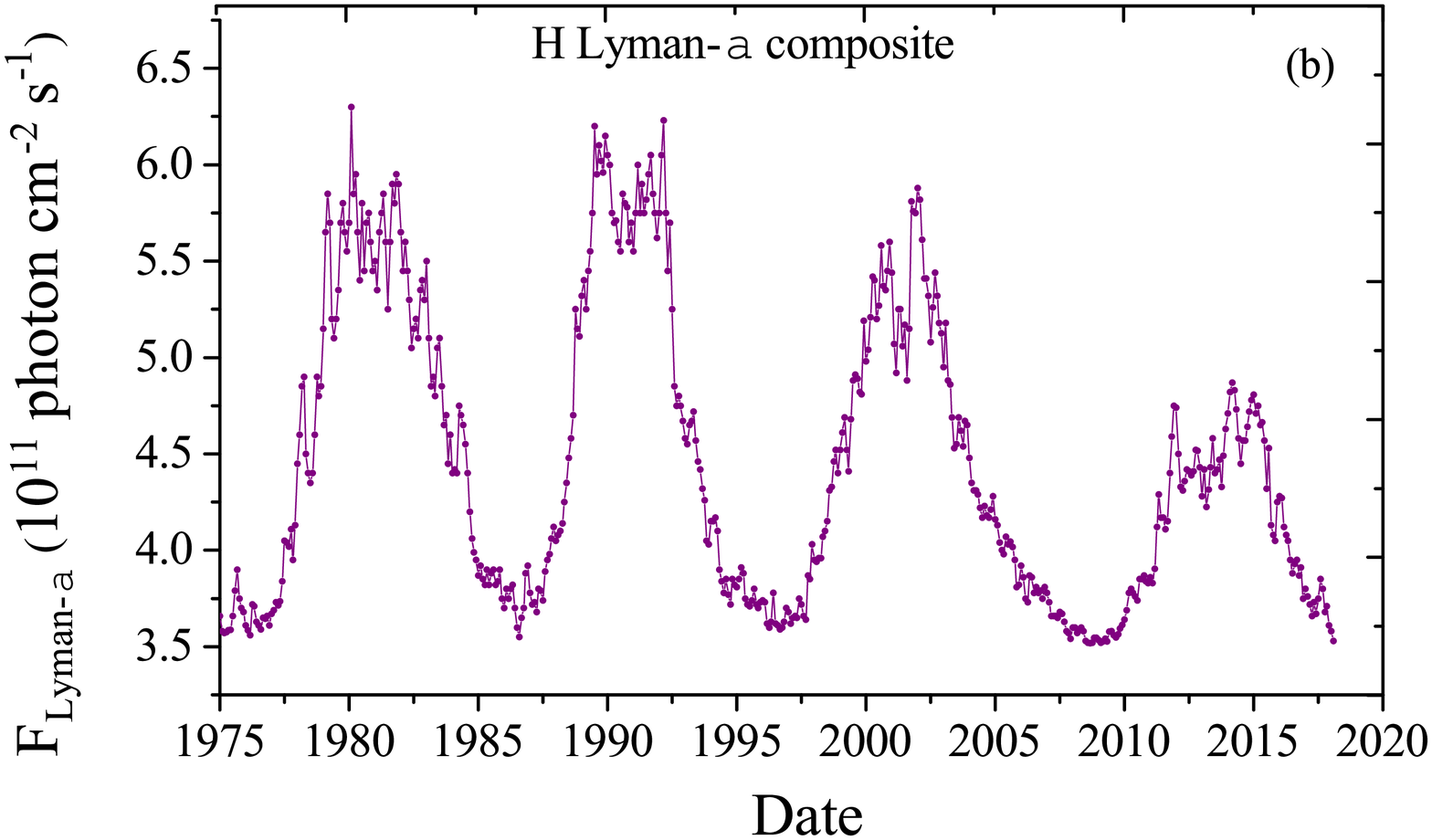}
              }
    \vspace{-0.05\textwidth}   
     \centerline{\hspace*{0.015\textwidth}
               \includegraphics[width=0.56\textwidth,clip=]{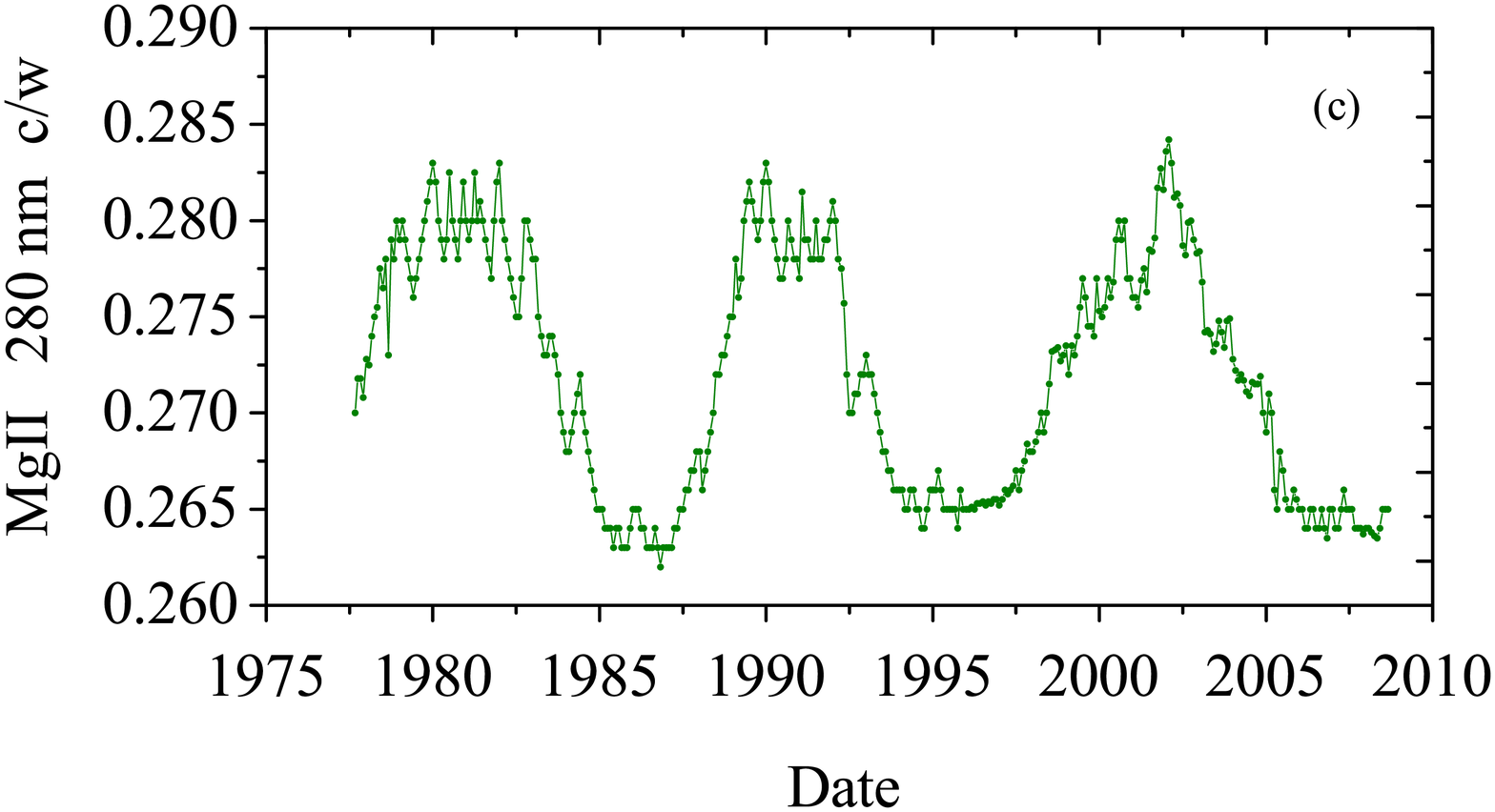}
               \hspace*{-0.01\textwidth}
               \includegraphics[width=0.57\textwidth,clip=]{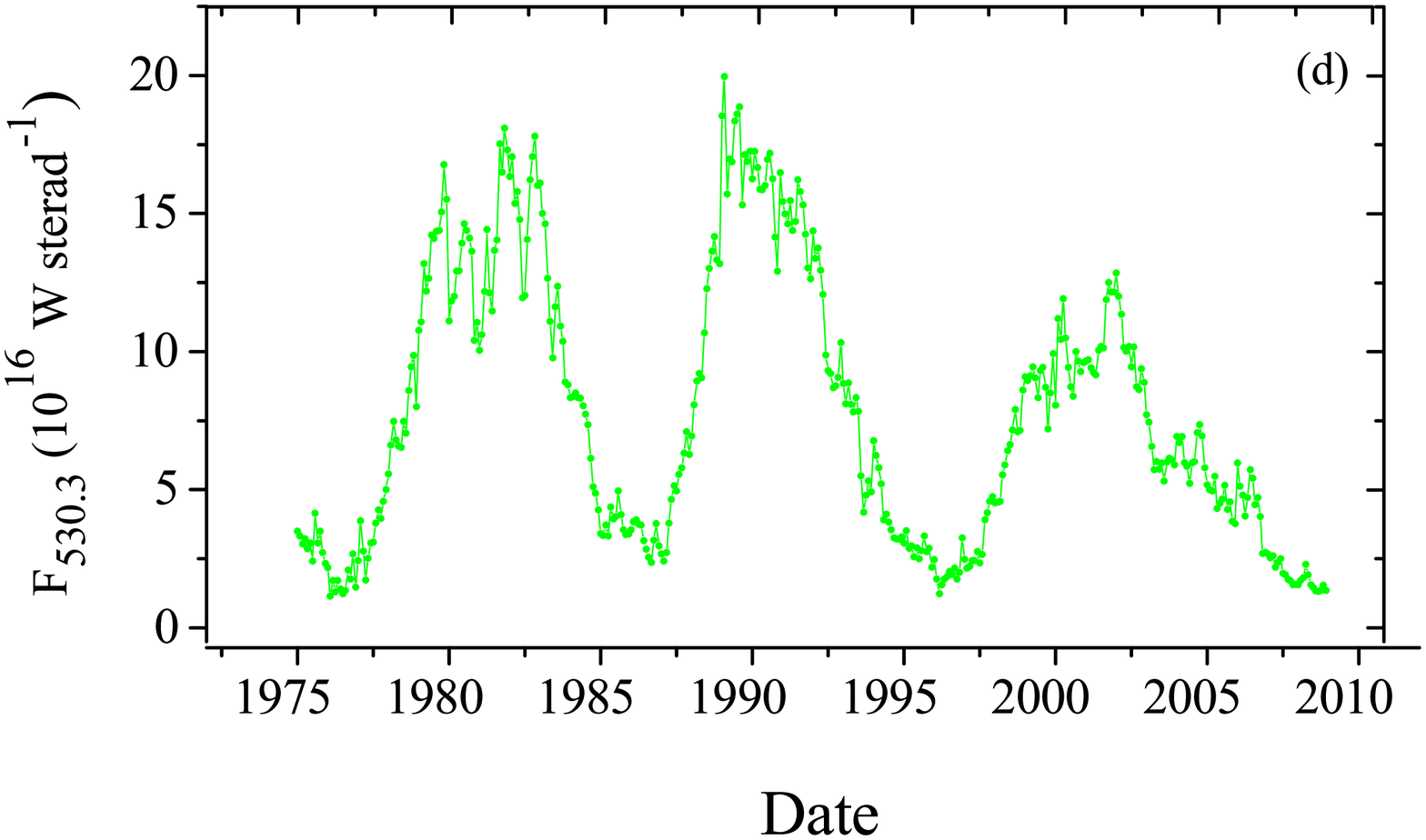}
              }
    \vspace{- 0.05\textwidth}   
    \centerline{\hspace*{0.015\textwidth}
               \includegraphics[width=0.56\textwidth,clip=]{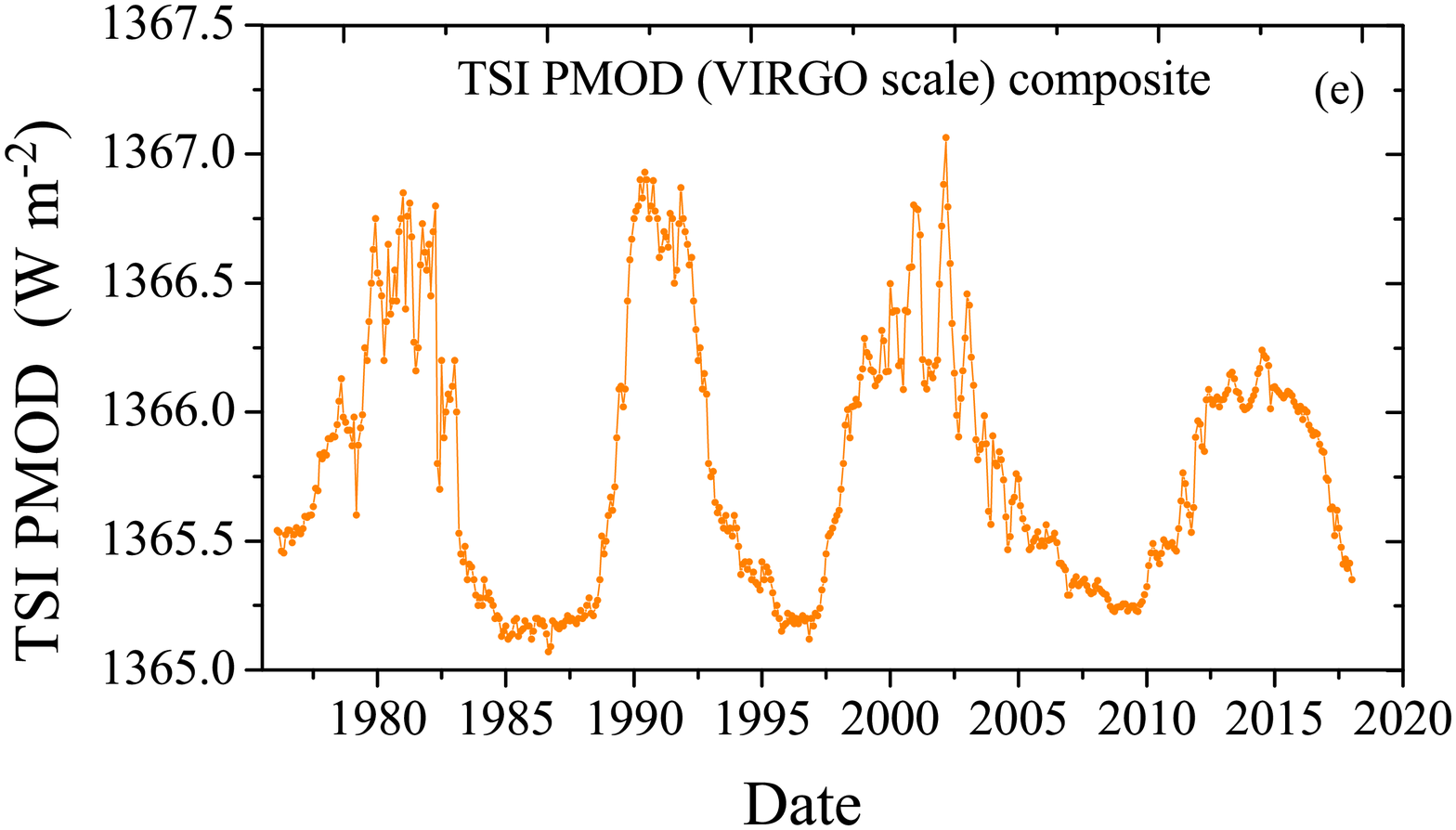}
               \hspace*{-0.03\textwidth}
               \includegraphics[width=0.53\textwidth,clip=]{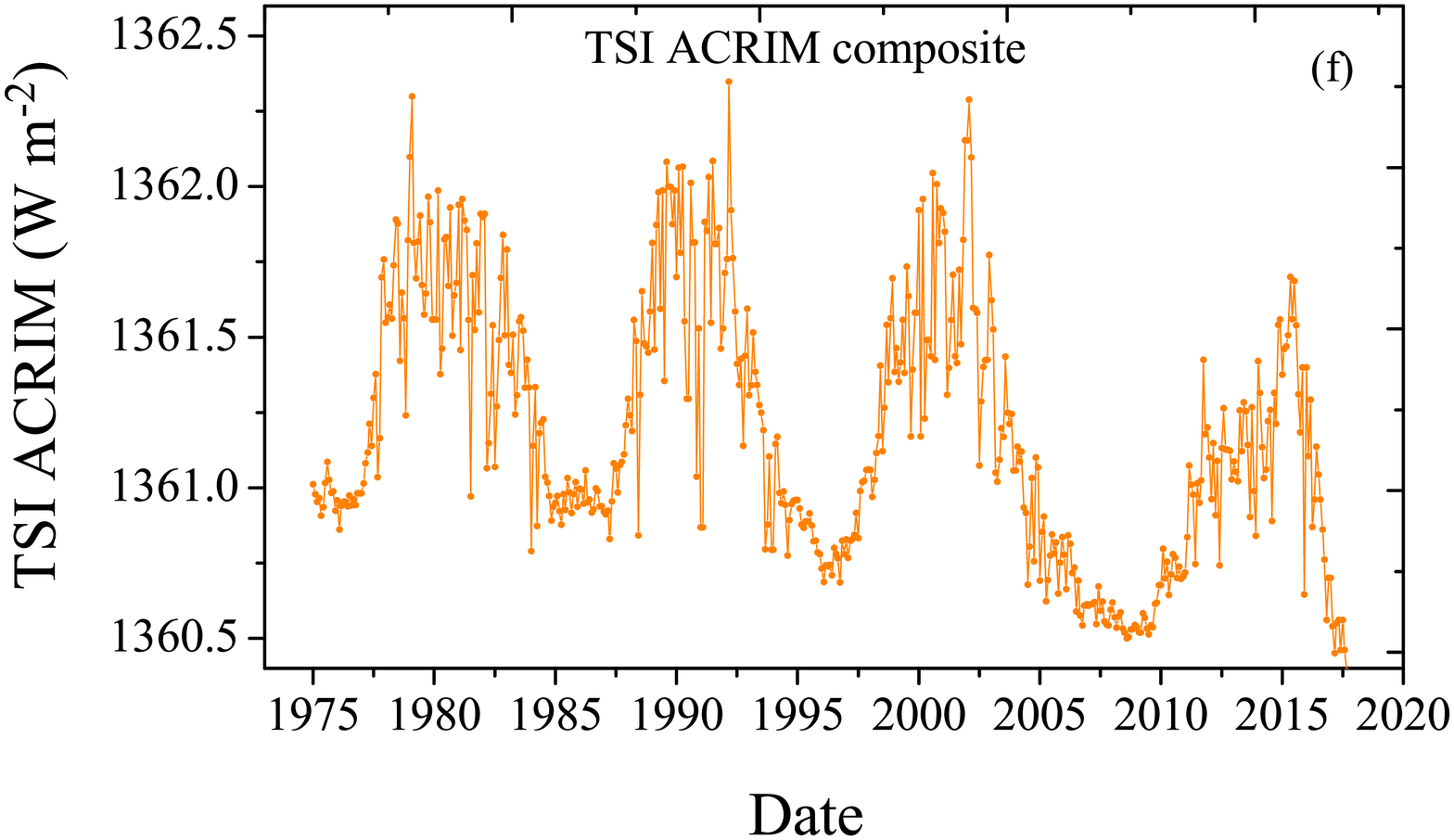}
              }
    \vspace{0.05\textwidth}   
    
     \vspace{-0.05\textwidth}    
       \centerline{\includegraphics[width=80mm]{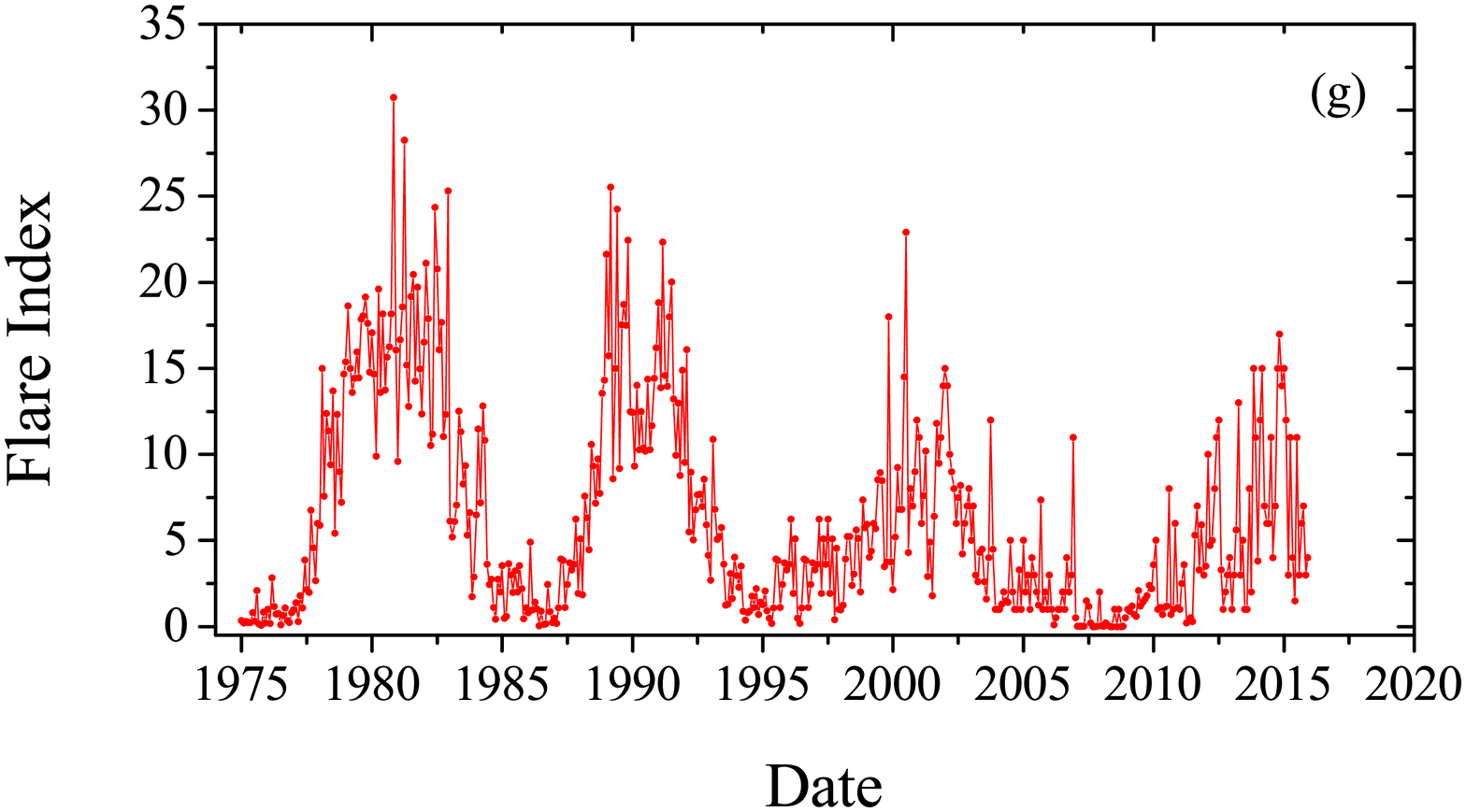}}
\caption{(a) International Sunspot Number (SSN),
the 1975 -- 2017 data;
~(b) H Lyman-$\alpha$ flux composite ($F_{Ly-\alpha}^{obs}$),
the 1975 -- 2017 data;
~(c) MgII 280 nm c/w Bremen composite, the 1975 -- 2017 data; 
~(d) $F_{530.3}$, the 1975 -- 2008 data;
~(e) TSI~PMOD composite (VIRGO scale), the 1978 -- 2017 data;
~(f) TSI ACRIM composite, the 1975 -- 2017 data;
~(g) The Flare Index, the (1975 -- 2017) data.}
  \vspace{0.05\textwidth}   
\end{figure}

Figure 2 shows the time series of observations of solar activity indices which are analysed in this work. For example in Figure 2a we show monthly SSN  for cycles 21 -- 24. The data was obtained from the available websites, see the information from the Table 1.

\begin{table}
\caption{Indices of solar activity and their observation data archives}
\begin{center}

\begin{tabular}{cccc}     
  \hline                   
 Activity Indices&Area of formation&Interval of observations& Web-sites with archive data    \\ \hline

  &  & & \\
 $F_{10.7}$  & corona &1975--2017& https://www. ngdc. noaa. gov/ stp/ \\
  &  & & solar/solardataservices.html. \\
SSN   & photosphere &1975--2017& http://www. ngdc. noaa. gov/ stp/ \\
&  & &sunspot-numbers/ international/. \\

$F_{Ly-\alpha}$ & chromosphere  &1975--2017 & http://lasp. colorado. edu/ lisird/ tss/ \\
&  & &composite\_lyman\_alpha.html.\\

MgII c/w & chromosphere  & 1975--2017& http://www.iup.uni-bremen.de/ \\
&  & & UVSAT/Datasets/mgii.\\

$F_{530.3}$ & corona &1975--2008 & https://www.ngdc.noaa.gov/stp/ \\
&  & &solar/corona.html \\

TSI~PMOD & photosphere &1978--2017 &https://www.pmodwrc.ch/pmod.php? \\
&  & &topic=tsi /composite/SolarConstant \\

TSI~ACRIM & photosphere &1975--2017 & ftp://ftp.ngdc.noaa.gov/STP/SOLAR\\
&  & &\_DATA/SOLAR\_IRRADIANCE/ACRIM3/\\

Flare Index & corona &1975--2016 & ftp://ftp. ngdc. noaa. gov/ STP/ \\
&  & &SOLAR DATA/ SOLAR FLARES/ INDEX. \\

\hline

\end{tabular}\label{tab1}
\end{center}
\end{table}

\section{The relationship of solar activity indices with of solar activity indices with $F_{10.7} $ in 1950 -- 1990} 
{\label{S:place}}

Tobiska et al. (2000)  pointed out that variations of the TSI,
solat Ultraviolet (UV), solar extreme ultraviolet (EUV) and solar soft X-rays
are the fundamental mechanisms causing variations in main parameters of the
terrestrial atmosphere, land and oceans.

First we have analyzed an interconnection between activity indices
SSN, MgII c/w (core to wing ratio), TSI composite (PMOD and ACRIM), H
Lyman-$\alpha$ 121.6 nm flux and Flare Index versus $F_{10.7}$  for
the 1975 -- 1990 data.

The $F_{10.7}$ index is the flux at the wavelength of 10.7 cm (2800 MHz) which comes from the full disc of the Sun. In comparison with the SSN the $F_{10.7}$ index has following advantages: it is the measure of real flux (so it is more objective) and it can be measured in any weather.

Note that in the period from 1950 to 1990 solar AI showed a
consistent and stable interrelations (Svalgaard 2013).
In Deng et al. (2013) the analyses of the phase asynchrony between 10.7 cm solar radio flux and sunspot numbers during the period of 1947 February to 2012 June
has done. It was shown that the phase asynchrony between coronal index and sunspot numbers has investigated. Deng et al. (2013) found that the sunspot numbers begin one month earlier than coronal index. This effect of asynchrony of AI versus $F_{10.7}$ (value of asynchrony is an order of month) introduces a small additional variance of values in the relations linking the activity Indices to the 10.7 flux studied in our paper, see Figure 4a (SSN versus $F_{10.7}$) and Figure 7a (coronal index versus $F_{10.7}$). On the other hand, the choice of $F_{10.7}$ to describe the average value of activity of the Sun seemed to us the most acceptable for objectives of the study of solar indices using normalized values, as is successfully done in (Svalgaard 2013). If we will take into account an effect of asynchrony between AI and 10.7 cm flux, then according to our estimates, this would slightly reduce the standard deviations in determining the regression coefficients. But the changes in the values of the regression coefficients and their errors (see Table 2) for our task are small enough. And when we calculate relative normalized indices AIFF, this effect generally becomes indistinguishable.

An interconnection between activity indices (in the period from 1950 to 1990 when solar AI showed a stable interrelations (Svalgaard 2013)) corresponds to the
polynomial regression equation:

\begin{equation}
     F_{ind}^{synt}  = a_{ind} + b1_{ind} \cdot F_{10.7}+ b2_{ind} \cdot F_{10.7}^2 \,.
\end{equation}

were  $F_{ind}^{synt}$ is the calculated AI flux,
$a_{ind}$, $b1_{ind}$ and $b2_{ind}$ are polynomial coefficients.

In the Table 2 the coefficients of polynomial
regression ($a$, $b1$, $b2$) for solar activity indices versus
$F_{10.7}$ in 1950~-- 1990 (when solar AI showed 
stable interrelations)  are presented . We have estimated the residual sum of squares
(RSS) both for linear and polynomial regressions of the AI versus the 
$F_{10.7}$. 
In this paper we use a polynomial regression for
$F_{ind}^{synt}$ calculations from equation (3).

Linear and polynomial regressions of the AI versus the $F_{10.7}$ are
shown in Figures 4a -- 10a below.

For different AI we calculated the AIFF which is equal to
$F_{ind}^{obs}/F_{ind}^{synt}$ according to the equation (1), where $F_{ind}^{synt}$ is calculated according to the equation (2).

Time dependencies of AIFF in the period from 1990 to the present 
show main trends in long-term changes of solar activity indices.

It's evident that the AIFF  remained close to 1 about until 1990,
but the  AIFF for different indices vary in different ways  from
1990 to the present, see Figure 4b -- 10b below.

\begin{table*}
\caption{Coefficients of polynomial regression from equation (3) for different AI versus $F_{10.7}$ for 1950 -- 1990 observations.}
\begin{center}
\begin{tabular}{ccccccc}     
  \hline                   
 $F_{ind}$ (AI) &$a$ &  $b1$  &  $b2$ &  $\sigma_a$ & $\sigma_{b1}$ &
 $\sigma_{b2}$\\ \hline
 
 & &  &  &   & & \\ 
SSN        & $-82.56$ &  1.39 &$-0.0011$   &  3.2 & 0.05 &  0.00016\\

$F_{Ly-\alpha}$& 2.01  & 0.0264 &$-4.3\cdot10^{-5}$  & 0.166 &0.0027  &$9.2\cdot10^{-6}$\\

MgII c/w & 0.134 & $2.7\cdot10^{-4}$  & $-5.46\cdot10^{-7} $& 0.0018&$2.8\cdot10^{-5}$  &  $9.55\cdot10^{-8}$\\

$F_{530.3}$& $-9.7$ & 0.203  &$-0.00041$ & 1.1  & 0.022   &  $7.6\cdot10^{-5}$\\

TSI~PMOD& 1364.5 & 0.01065  &$-7.13\cdot10^{-6}$&0.198  &
0.001&$0.8\cdot10^{-7}$\\

TSI~ACRIM& 1359.9 & 0.0179 &$-4.6\cdot10^{-5}$& 0.137  & 0.0022   &$7.7\cdot10^{-6}$\\

Flare Index &$-1.358$ & 0.0019 & 0.00042 & 0.046   &  0.00015  &$8.8\cdot10^{-5}$\\

\hline

\end{tabular}\label{tab2}
\end{center}
\end{table*}

\subsection{ The $F_{10.7}$ global index.}

The $F_{10.7}$ index is most suitable index for solar activity estimation and forecasting.
$F_{10.7}$ index along with SSN has one of longest running records of solar
activity. This very popular global index is measured in present time in Penticton, British Columbia.

The $F_{10.7}$  is an useful indicator  of the solar atmosphere's
emissions which is created by radiant solar active regions and whose
energies are very important for the formation of the thermosphere of
the Earth. The  EUV -- $F_{10.7}$ close correlation
are often use in many Earth's atmospheric models (Tobiska et al. 2008).

It was shown that the 10.7 cm emission can be separated into next  components: temporal events which are caused by flares with about an hour duration, long-term variations  with durations up to years and the constant component of the minimum value -- Quiet Sun
Level (Tobiska et al. 2008). 
Donnelly et al. (1983) pointed out a good correlation of
$F_{10.7}$ with full-disc flux in CaII and MgII lines. 
EUV solar emissions act to Earth's ionosphere and modify it.
 These emissions correlate well with the $F_{10.7}$. 
UV emissions which influence to the stratosphere of the Earth are also
correlate with the $F_{10.7}$ index (Chapman \& Neupert 1974; 
Bruevich \& Nusinov 1984; Lean 1987). 

\begin{figure}    
   \centerline{\includegraphics[width=80mm]{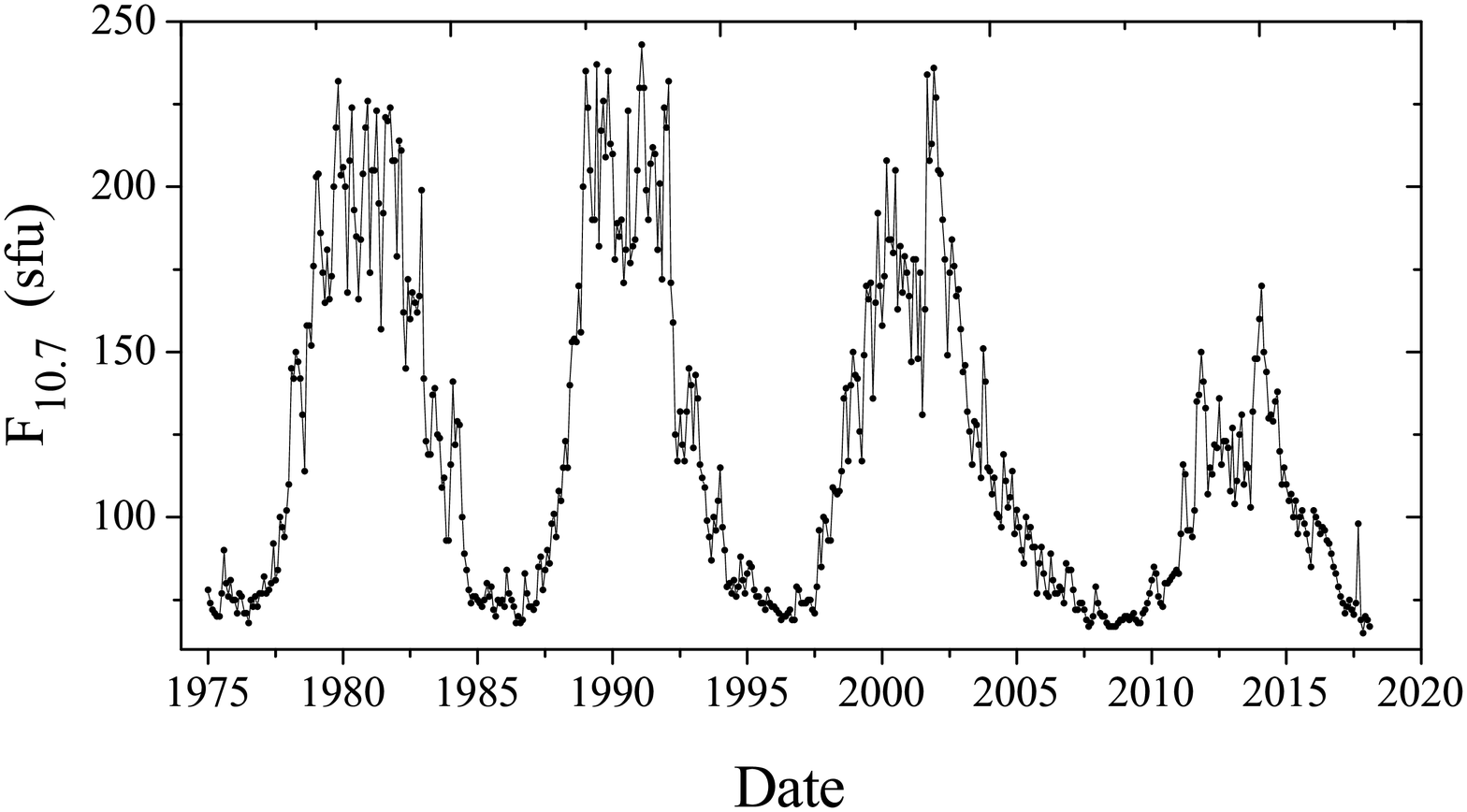}}
\caption{The $F_{10.7}$ global index. Observations for 1975 -- 2017 }
\end{figure}

Unlike many solar indices, the  $F_{10.7}$ radio flux can easily be
measured reliably on a day-to-day basis from the Earth's surface, in
all types of weather. 
Reported in "solar flux units" (sfu), the
$F_{10.7}$ can vary from  67 -- 68 sfu, to above 300 sfu in dependence of a 
phase of a solar cycle.

We used NASA data of $F_{10.7}$ index from National Geophysical Data Center (NGDC) web-site, see Table 1. 

In Figure 3 we can see the $F_{10.7}$ monthly data for cycles 21 – 24.
It is evident that for the
$F_{10.7}$  index as in the case of SSN in Figure 1 the maximum of the
amplitude in cycle 24 is approximatively two times less than in 
cycles 21 -- 23.

\subsection{SSN}

Sunspots are temporary events on the Sun's photosphere.  
We can see sunspots as dark areas compared to surrounding surface.

 Solar spots are places with strong concentrations of local magnetic fields. These powerful magnetic fields inhibit convection, so spots on solar surface are regions of lowered temperature. The series of direct SSN observations
continue more than two hundred years.

R. Wolf  in 1850  has defined the relative sunspot number as $R = 10 \cdot Number ~of ~Groups + Number ~of ~Spots $ ~which an observer can see on the solar disk.

Svalgaard \& Cliver (2010) has recognized that most of solar AI are close correlated with SSN. In our paper we use the SSN data from NASA NGDC Web-site, see  Table 1. 

In Figure 4a we show monthly  SSN versus $F_{10.7}$ for the period of stable interconnections between solar indices (1950 -- 1990). The linear regression correlation coefficient for SSN versus $F_{10.7}$ is high (equal to 0.96).
Using the time dependence of the SFF -- $SSN^{obs}/ SSN^{synt} $ we have
analyzed long-term trends in normalized variations of the
SSN-index.

 We have also estimated RSS both for the linear and for the
polynomial regression of the SSN versus $F_{10.7}$ (Figure 4b). The
RSS has the the lower value for the polynomial regression. So, a polynomial regression  best describes the temporal variation in $SSN^{obs}/ SSN^{synt} $.

According to the trend which is displayed by the linear regression,
the normalized SSN-index ($SSN^{obs}/ SSN^{synt} $ in Figure 4b)
steadily decreases from 1950 to the present time approximately from
1 to 0.8.

We assume that the tendency which is displayed by the polynomial
regression describes more than a short-time trend. The polynomial
trend shows that the normalized SSN-index $SSN^{obs}/ SSN^{synt} $ has
stayed almost constant from 1950 to 1990 and has been equal to 1. But in
the period from 1990 to 2017 the $SSN^{obs}/ SSN^{synt} $ changed
strongly,  from 1 to 0.7 which is correspond to SSN study 
 in Nagovitsyn et al. (2012). Thus, polynomial fit of normalized SSN data shows that there has been a sharp decrease in the normalized SSN-index of sunspots $SSN^{obs}/ SSN^{synt} $ in the recent time (mainly in 23 and 24 cycles).

\begin{figure}    

   \centerline{\hspace*{0.015\textwidth}
               \includegraphics[width=0.55\textwidth,clip=]{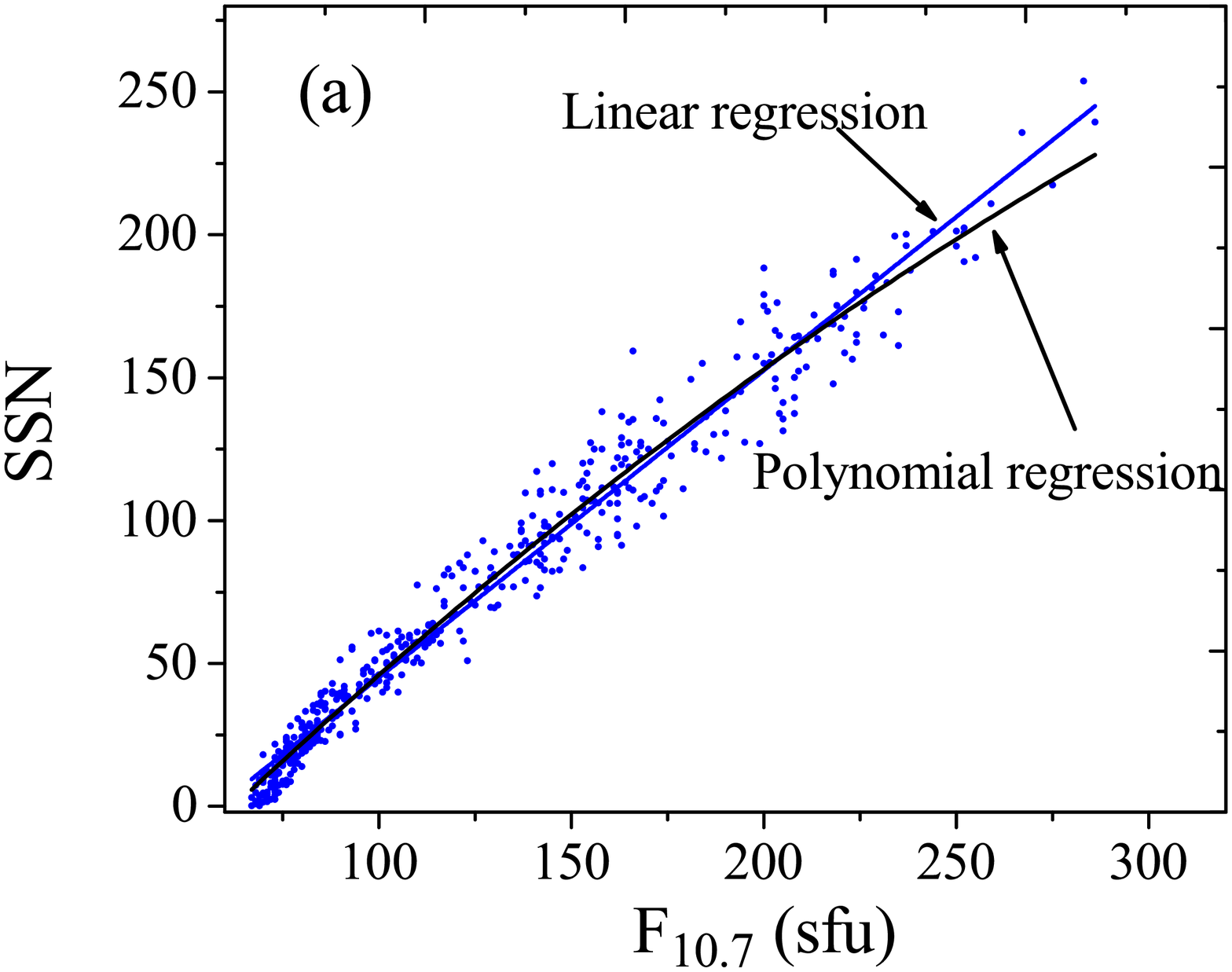}
               \hspace*{-0.015\textwidth}
               \includegraphics[width=0.55\textwidth,clip=]{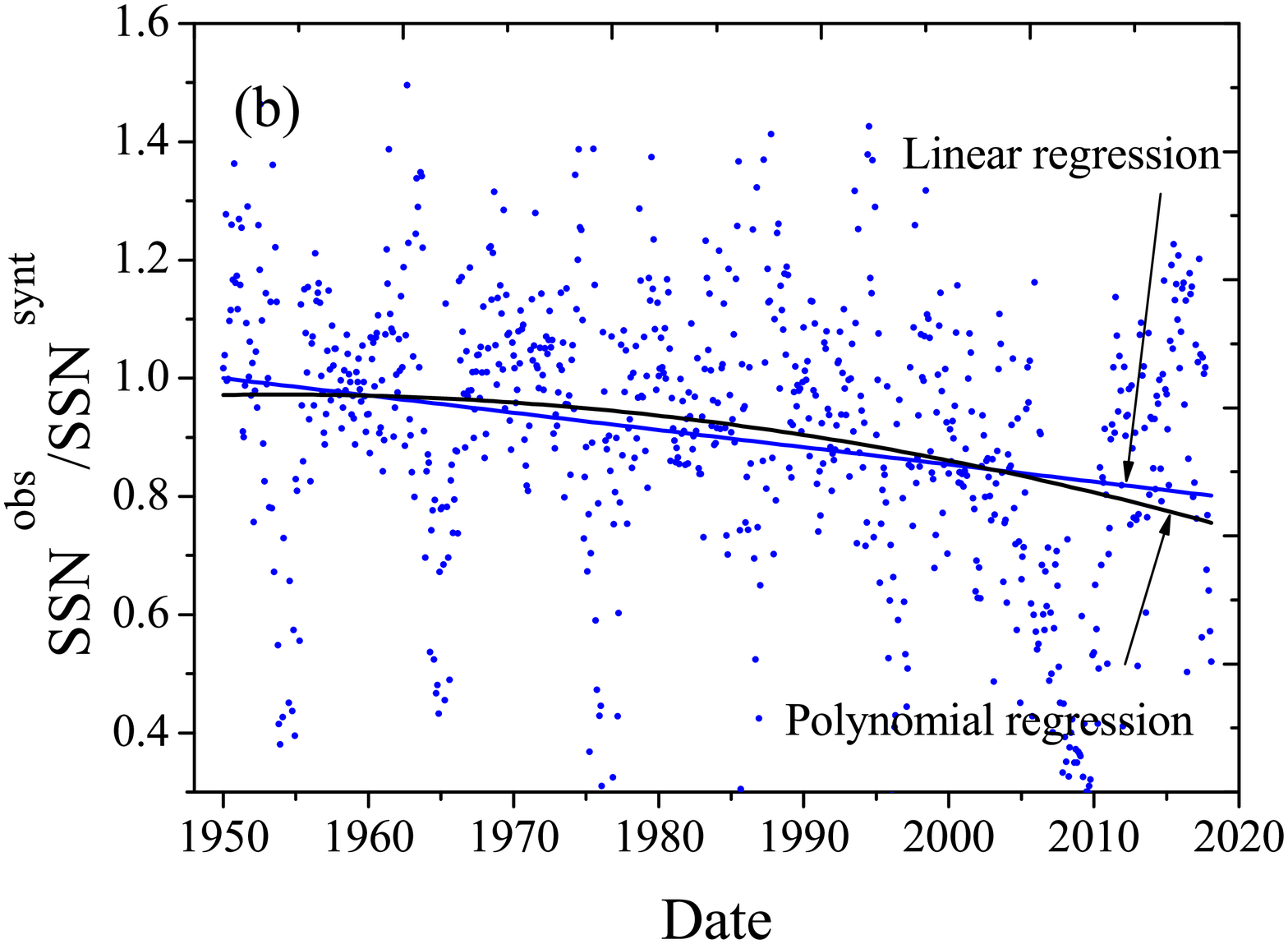}
              }
    \vspace{-0.15\textwidth}   
     \vspace{0.15\textwidth}    
\caption{(a) SSN versus $F_{10.7}$ for the period from 1950 to 1990;
 (b)AIFF for SSN (SFF) -- $SSN^{obs}/ SSN^{synt} $ for the period from 1950 to 2017.
        }
   \label{F-3panels}
\end{figure}

\subsection{Ultraviolet solar emission}

Solar radiation in ultraviolet (UV) range, which varies over several temporal scales (11-yr, 22-yr, 25 d, etc.), modulates remarkably the evolution of the ionosphere and thermosphere of the Earth. 
The solar activity dependence of the ionosphere is a key and fundamental issue in ionospheric physics, providing information essential to understanding  the  variations  in  the  ionosphere and  its  processes.  Solar UV-fluxes variations are important to research of the earth's climatic impact of solar variability. 
More hard solar emissions -- EUV (Extreme UV) and X-rays -- is the main source of perturbation in the upper atmosphere of the Earth.
The chromospheric  density and temperature in the Sun 
determine the intensities of the CaII and MgII spectral lines, the
coronal density and temperature produce the EUV and soft X-rays.

The UV-flux affects  to the
troposphere, to the photo dissociate process of the major atmospheric
constituents in the stratosphere, heats the stratosphere,
which may influence tropospheric as well as stratospheric dynamical
processes (Donnelly et al. 1983).

 The F2 peak in Earth's ionosphere which is characterized by the maximum value of electron density has a close correlation with solar activity in EUV (Liu et al. 2011). Solar short-wave flux is also affect the critical frequency of the F2 layer (foF2), the plasma temperature and thermosphere winds and other important for climatic forecasts parameters of upper atmosphere of the Earth. The best relationship is shown by the fluxes in the EUV and MgII 280 nm c/w index.

The variability in solar EUV-UV radiation during the solar cycle affects the state of the upper atmosphere of the Earth more noticeably than the more powerful radiation in the optical range that passes through the ionosphere, practically without affecting it.

EUV-UV emission lines dominate significantly over the emission in continuum part of solar spectrum. These emissions are formed under conditions of non-local thermodynamic equilibrium. Places of their formation are the layers of solar
atmosphere with high temperature (transition region and low corona).

The shorter the range of emission -- the higher the amplitude of long-term variability: the change during solar cycle reaches factors of 2 for irradiance in 10 -- 120 nm range, then reaches 1-10\% for 120 -- 400 nm irradiance and change in solar cycle is less than 1\% for the TSI
 (Lean 1987; Tobiska et al. 2000).

Emission at 121.6 nm (H Lyman-$\alpha$ ) affects mainly the mesosphere of the Earth. H Lyman-$\alpha$ photons photoionizes NO molecules and form D-region of ionosphere.  H Lyman-$\alpha$ emission is also actively participate in the process of water vapor dissipation and after this active hydroxyl radicals are produced.

The effect of HeII 30.4 nm photons on heating of Earth's thermosphere is the most important.
The flux in HeII 30.4 nm line correlates most closely with the fluxes in lines of the
ultraviolet part of the solar spectrum (Bruevich \& Nusinov 1984). In
this regard, the study of the flux in the Lyman-alpha line
$F_{Ly-\alpha}$ and its cyclic variations is a very important task
for the  forecasting of the state of the Earth's thermosphere.

In our paper we use the data set of Composite Solar Lyman-alpha which was made in  University
of Colorado. These data are available at Web-site, see Table 1. 

These data are based on observations from satellites TIMED, SOURCE
combined with earlier missions SME, AE-E, NOAA, all the data were
calibrated to the UARS (Upper Atmosphere Research Satellite)
SOLSTICE level.

In Figure 2b we see H Lyman-$\alpha$ flux composite observed data --
($F_{Ly-\alpha}^{obs}$). It is seen that the absolute values of
the $F_{Ly-\alpha}^{obs}$ in the cycle  24 are about 30 \% less than
in the cycles 21 -- 23. 

In Figure 5a we show monthly
$F_{Ly-\alpha}^{obs}$ versus $F_{10.7}$ for the period of stable
interconnections between solar indices 1950 -- 1990. The linear regression correlation coefficient is high (equal to 0.94).
It is almost also very high as in the case of dependencies on the Figure 4a.

The time dependence of AIFF for H Lyman-$\alpha$ irradiance  
-- $F_{Ly-\alpha}^{obs}/ F_{Ly-\alpha}^{synt}$ (LalFF)
in Figure 5b shows the long-term trend in the normalized variation of
the $F_{Ly-\alpha}$-index. According to the trends which are
displayed both by linear and polynomial  regressions we see that the
normalized $F_{Ly-\alpha}$-index (LalFF) remains constant (with an accuracy of 2-3 \%)and is about 1 in the recent time.

\begin{figure}    
   \centerline{\hspace*{0.015\textwidth}
               \includegraphics[width=0.55\textwidth,clip=]{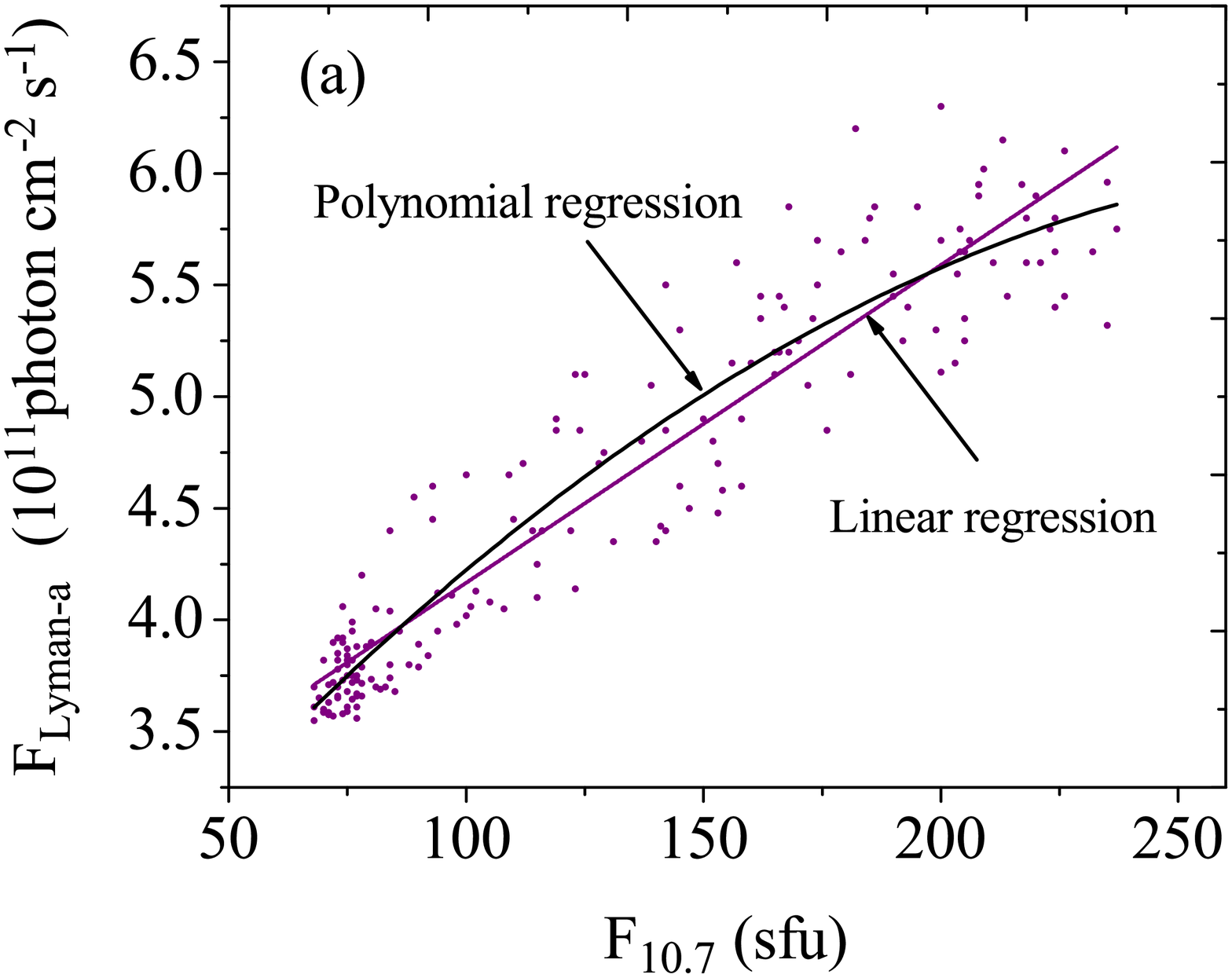}
               \hspace*{-0.015\textwidth}
               \includegraphics[width=0.55\textwidth,clip=]{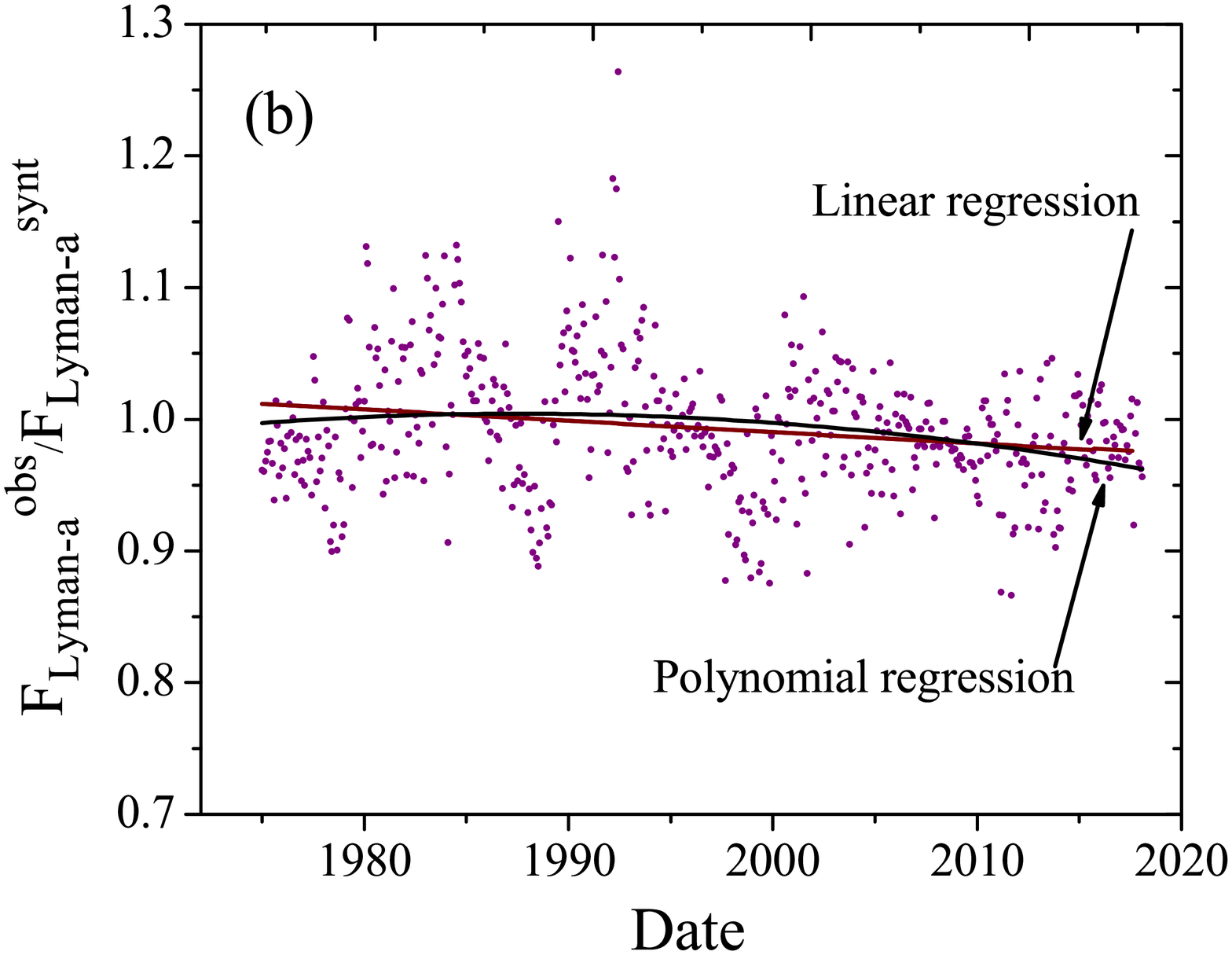}
              }
     \vspace{-0.35\textwidth}   
     \vspace{0.40\textwidth}    
\caption{(a) $F_{Ly-\alpha}^{obs}$ versus $F_{10.7}$ for the period from 1970 to 1990;
 (b) AIFF for H Lyman-$\alpha$ irradiance (LalFF) -- $F_{Ly-\alpha}^{obs}/ F_{Ly-\alpha}^{synt}$ for the period from 1970 to 2017.
        }
   \label{F-2panels}
\end{figure}

MgII c/w index is not quite unusual among other indexes. This index includes information about both the solar chromosphere (cores of MgII lines are located at 279.56 and 280.27nm)  and about solar photosphere (wings of MgII lines).

\begin{figure}    
   \centerline{\hspace*{0.015\textwidth}
               \includegraphics[width=0.56\textwidth,clip=]{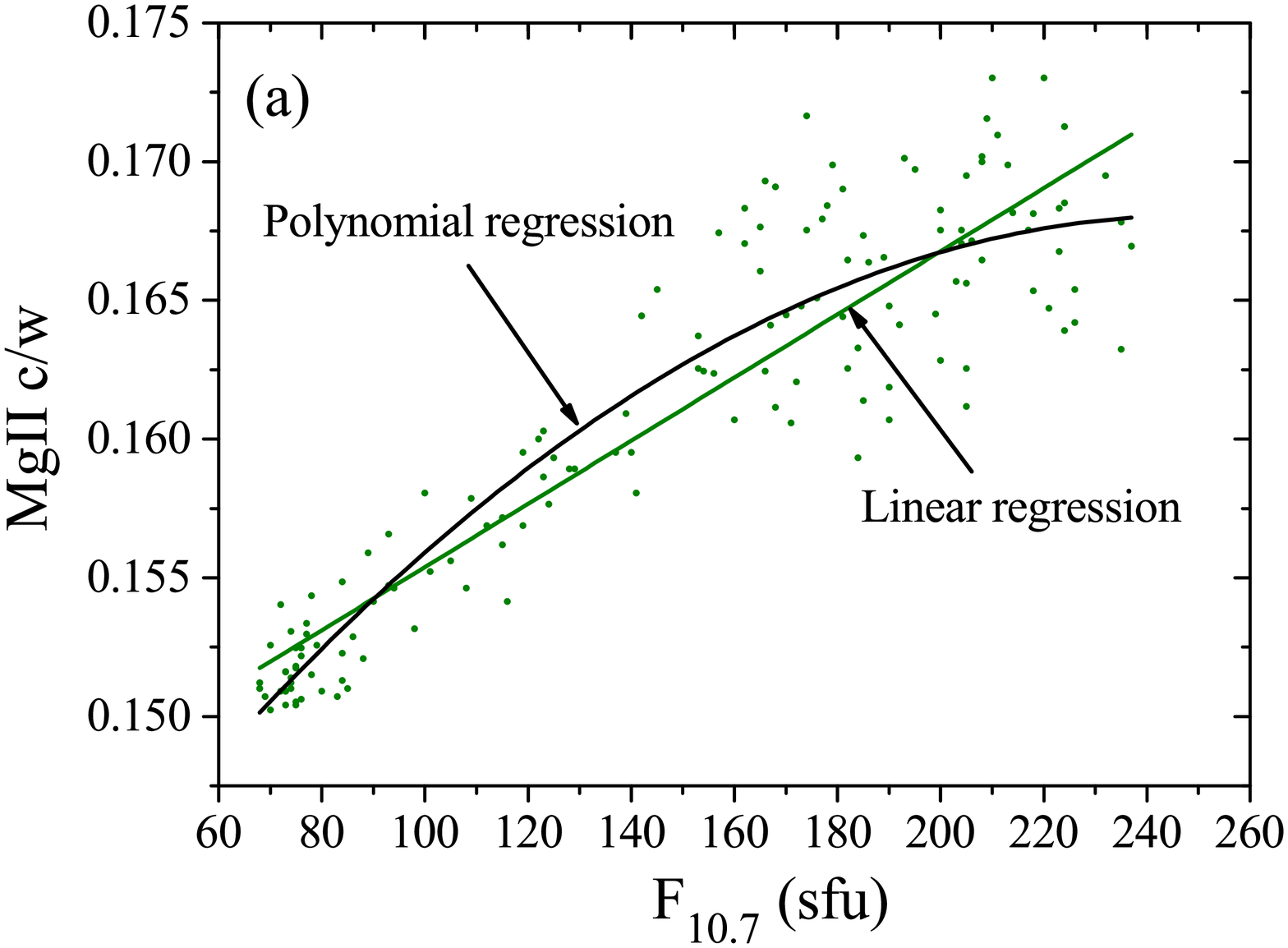}
               \hspace*{-0.015\textwidth}
               \includegraphics[width=0.56\textwidth,clip=]{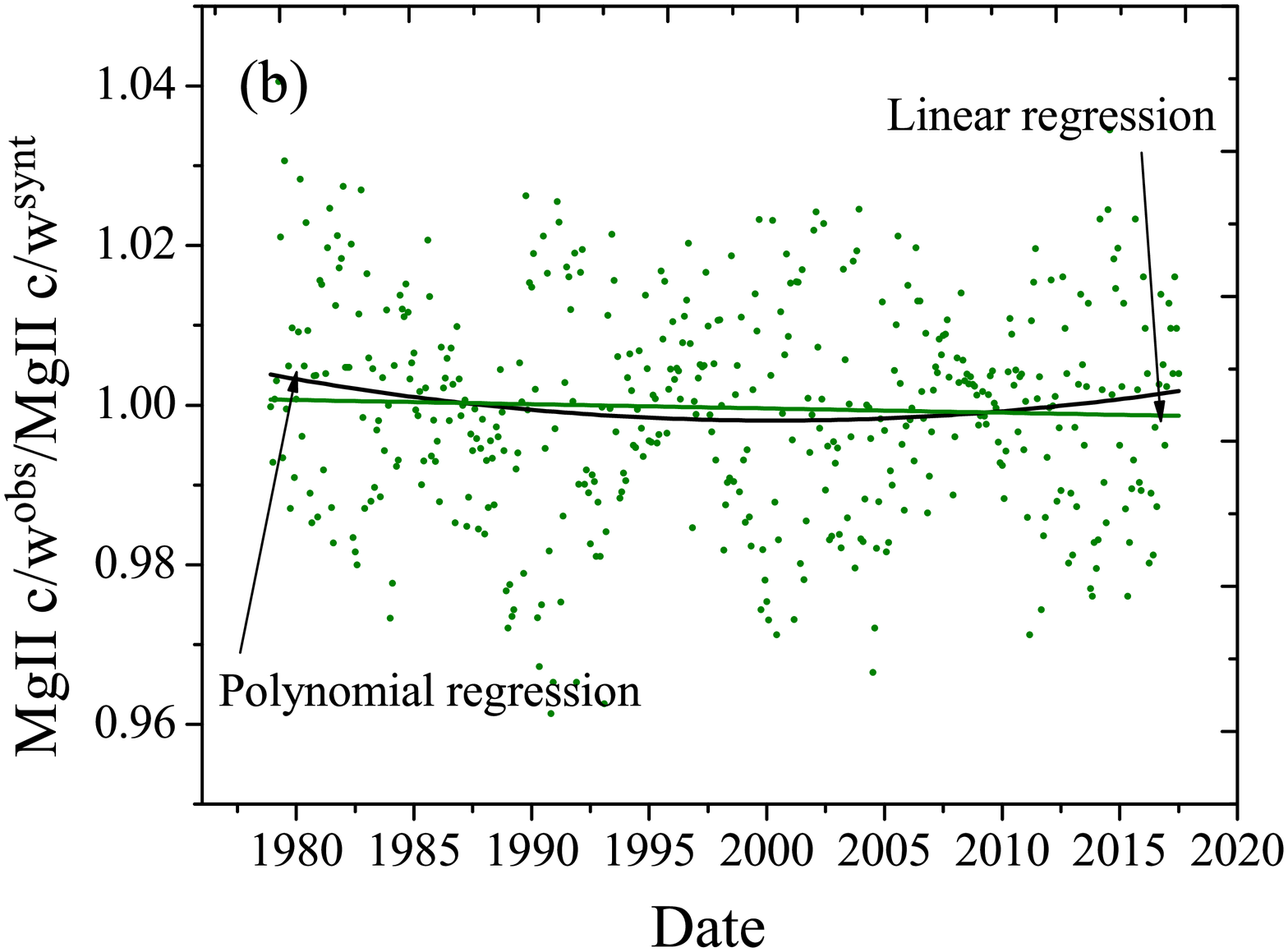}
              }
     \vspace{-0.35\textwidth}   
     \vspace{0.40\textwidth}    
\caption{(a) $ MgII ~c/w^{obs}$ versus $F_{10.7}$ for the period from 1978 to 1990;
 (b) AIFF for MgII 280 nm c/w irradiance (MgIIFF) -- $MgII ~c/w^{obs}/ MgII ~c/w^{synt}$ for the period from 1978 to 2017.
        }

   \label{F-2panels}
\end{figure}

Solar  MgII c/w is used for evaluation of common level  of solar activity (Floyd et al. 2005). MgII c/w -index has also the close correlation with other UV-EUV indices (Viereck \& Puga 2001).

NOAA started in 1978, ENVISAT was launched on 2002. The NOAA and
ENVISAT MgII c/w observation data are in a good agreement with 
(Skupin et al. 2005). We used the long time series data set from NOAA,
GOME, SCIAMACHY, and GOME-2 which have been combined into a single
composite time series from the University of Bremen, see Table 1.

For convenience, this composite will be referred to as the Bremen
composite. In addition to scaling the datasets, there have been a
number of additional corrections to the observations for this
composite (Snow et al. 2014).

In Figure 2c we show MgII c/w Bremen composite data.
It is seen that the in the cycles 21 - 23 the maximum amplitudes of
MgII c/w  was about 15 \% exceeding of the average value but in the cycle  24
the maximum amplitudes of MgII c/w  was about 8 -- 10 \% exceeding of the average. 

In Figure 6a we show monthly MgII c/w  versus $F_{10.7}$ for the period 1950 --
1990. The correlation coefficient of the linear regression is very
high too and is equal to 0.92.

The time dependence of normalized MgII c/w index (MgIIFF) -- 

$MgII ~c/w^{obs}/MgII ~c/w^{synt}$ in Figure 6b shows the trends which are displayed both by linear and polynomial
regressions. We see that the normalized MgII c/w-index remains approximately constant
and is about 1 in recent time. In comparison with the normalized
$F_{Ly-\alpha}$-index , which shows a very weak downward trend the
normalized MgII c/w-index shows a very weak increase trend.

\subsection{Coronal flux at a wavelength of 530.3 nm}

\begin{figure}    
 \centerline{\hspace*{0.015\textwidth}
               \includegraphics[width=0.53\textwidth,clip=]{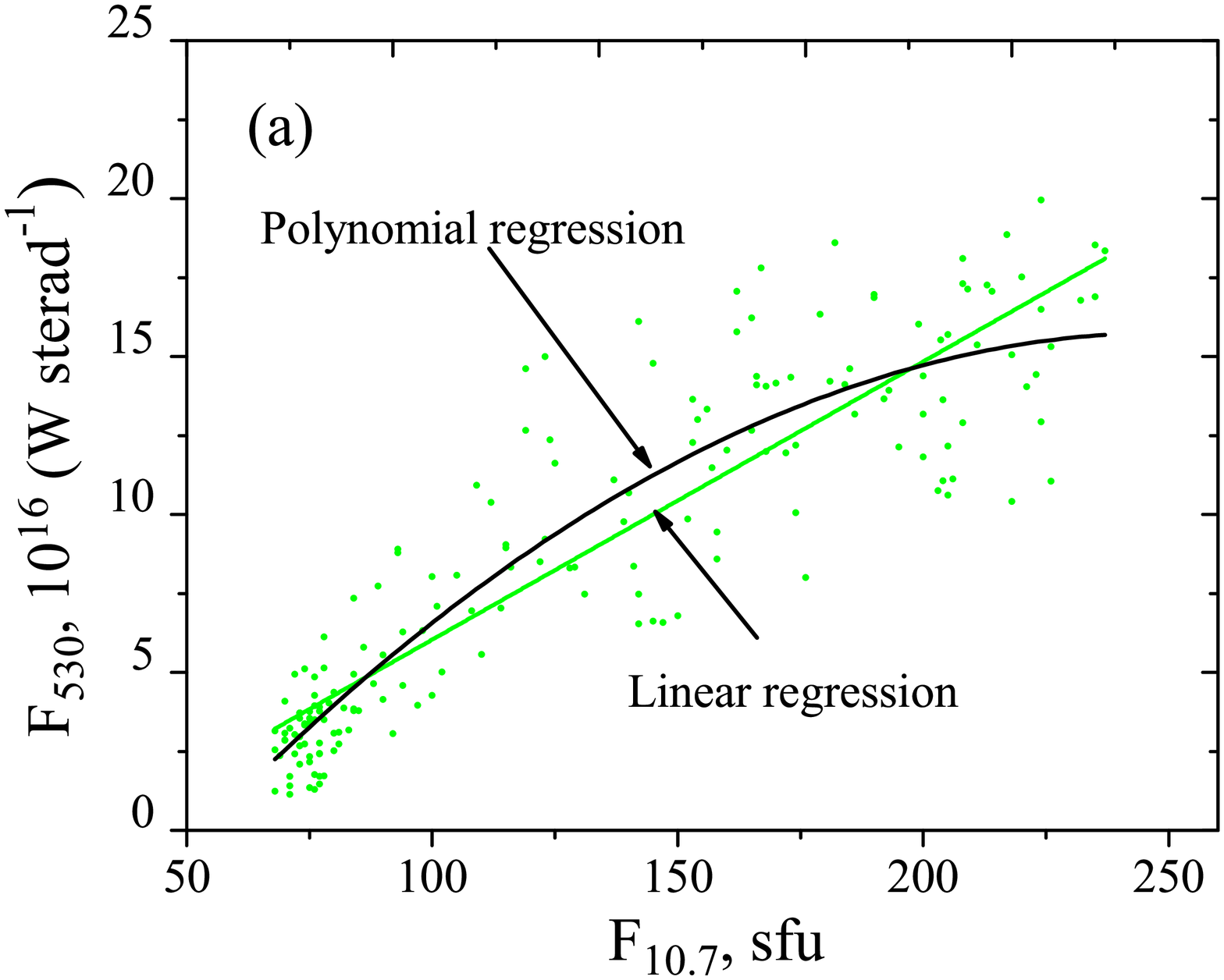}
               \hspace*{-0.015\textwidth}
               \includegraphics[width=0.535\textwidth,clip=]{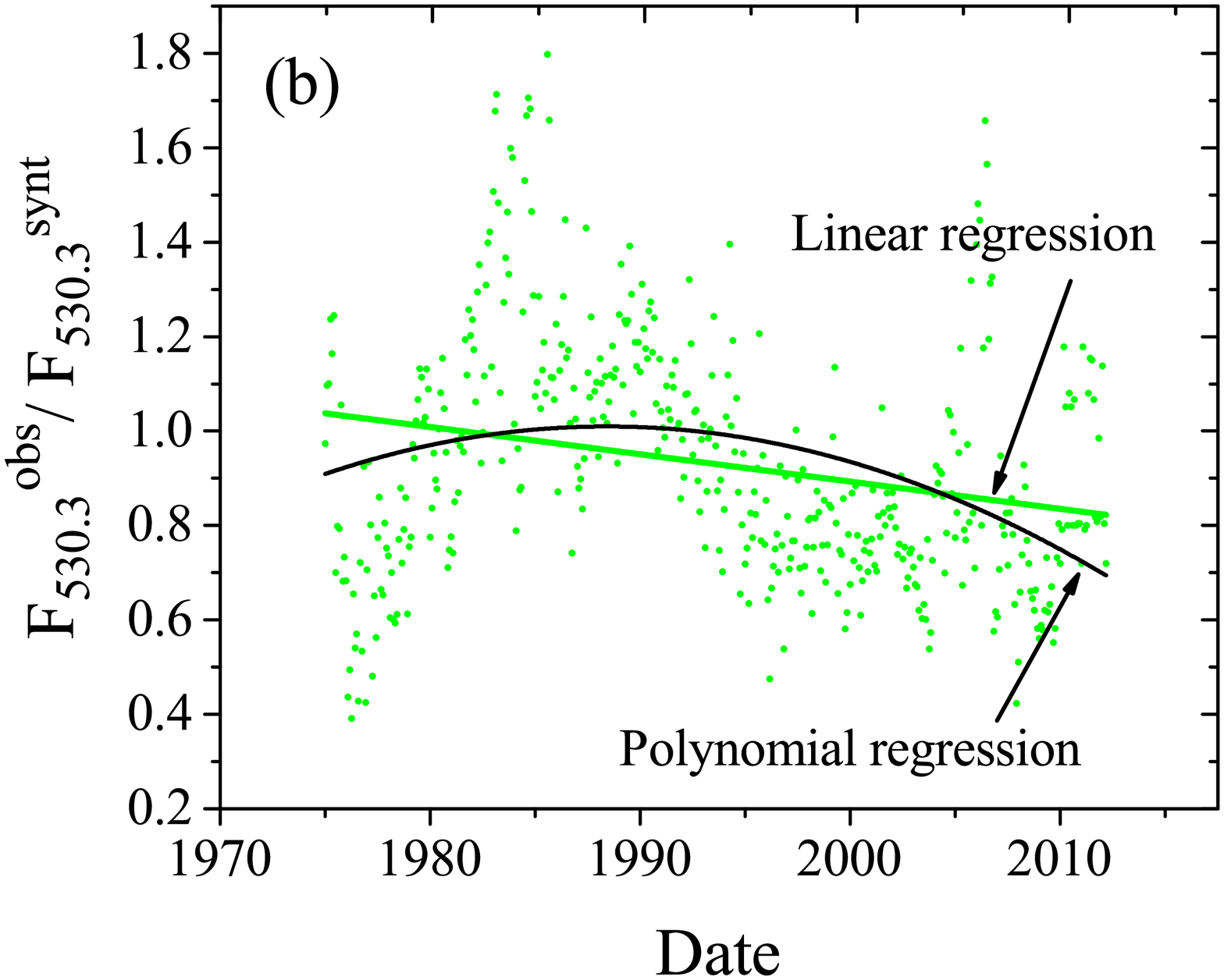}
              }
     \vspace{-0.35\textwidth}   
     \vspace{0.40\textwidth}    
\caption{(a)$F_{530.3}^{obs}$  versus $F_{10.7}$ for the period from 1950 to 1990;
 (b) AIFF for 530.3 nm irradiance (530FF) -- $F_{530.3}^{obs}/ F_{530.3}^{synt} $ for the period from 1975 to 2008.
        }
   \label{F-2panels}
\end{figure}

Solar flux at 530.3 nm ($F_{530.3}$) emitted by a green corona (FeXIV, 530.3 nm) is  much used index of solar activity  was introduced as so-called coronal index by (Rybansky 1975).  Variations of this index are associated with an evolution in the corona of bright structures which are characterized by different times of existence.
 Datasets of 530.3 nm flux are available for 1975 – 2008 at NGDC web-site, see Table 1.

In Deng et al. (2015) it was studied the interconnection of coronal index with sunspot-depended solar AI and it was claimed that the $F_{530.3}$ can depend on variations both of the SSN and solar flares. So that  the $F_{530.3}$ can be more useful in learning of solar-terrestrial connections.

The $F_{530.3}$  variations depend on the phase of solar cycle. These variations are in close connection with solar magnetic flux variations.

In Figure 2d we see datasets of monthly $F_{530.3}$.

It is seen that in cycles 21 -- 22 maximum amplitudes of the 
$F_{530.3}$ were approximately equal in both cycles but in cycle  23
the maximum amplitude of the $F_{530.3}$ was about 40 -- 50 \% less. 

In Figure 7a we show monthly $F_{530.3}$ versus the $F_{10.7}$ for the period 1950 -- 1990. The correlation coefficient of the linear regression in this case is
high enough and is equal to 0.90.

A time dependence of normalized $F_{530.3}$ (530FF) in Figure 7b shows a long-term negative 
trend in relative variations of the coronal 530.3 nm flux. According
to the trends which are displayed both by linear and polynomial
regressions we see that the 530FF shows a significant decreasing by about 20\% in cycle 23  as compared with previous two cycles.
In comparison with other solar activity indices (which have AIFF values approximately stable)
normalized  $F_{530.3}$  index 530FF shows a remarkable decrease in values.

\subsection{Total solar irradiance}

Total Solar Irradiance (TSI), known as Solar Constant, is most important index of solar activity which is the total flux of energy of Sun's radiation incoming to upper part of the Earth's atmosphere. Regular observations of the TSI on
satellites have being carried out since 1978 to the present.

A study of reliable datasets of the TSI  is very important for understanding of the problem of Earth's climate. It is very useful to know what contribution in the warming of the climate is made by TSI in comparison with industrial gases which are thrown in the atmosphere.

Krivova et al. (2003) showed that approximately to 63\% of changes in TSI is generated in the UV-EUV range.

On average TSI varies about 0.1\% in 11-yr cycle only. ACRIM, PMOD and other science
teams have been developed the physics-based models of TSI variations
depending on solar activity (Krivova \& Solanki 2008); Kopp et al. 2016). Two opposing processes make a contribution in TSI variability. These processes are a darkening due to sunspots and a  brightening due to faculae.

We will see below how long-time trends in the TSI coincide with
long-time trends in fluxes in UV/EUV and in other solar AI.

\begin{figure}    
   \centerline{\hspace*{0.015\textwidth}
               \includegraphics[width=0.55\textwidth,clip=]{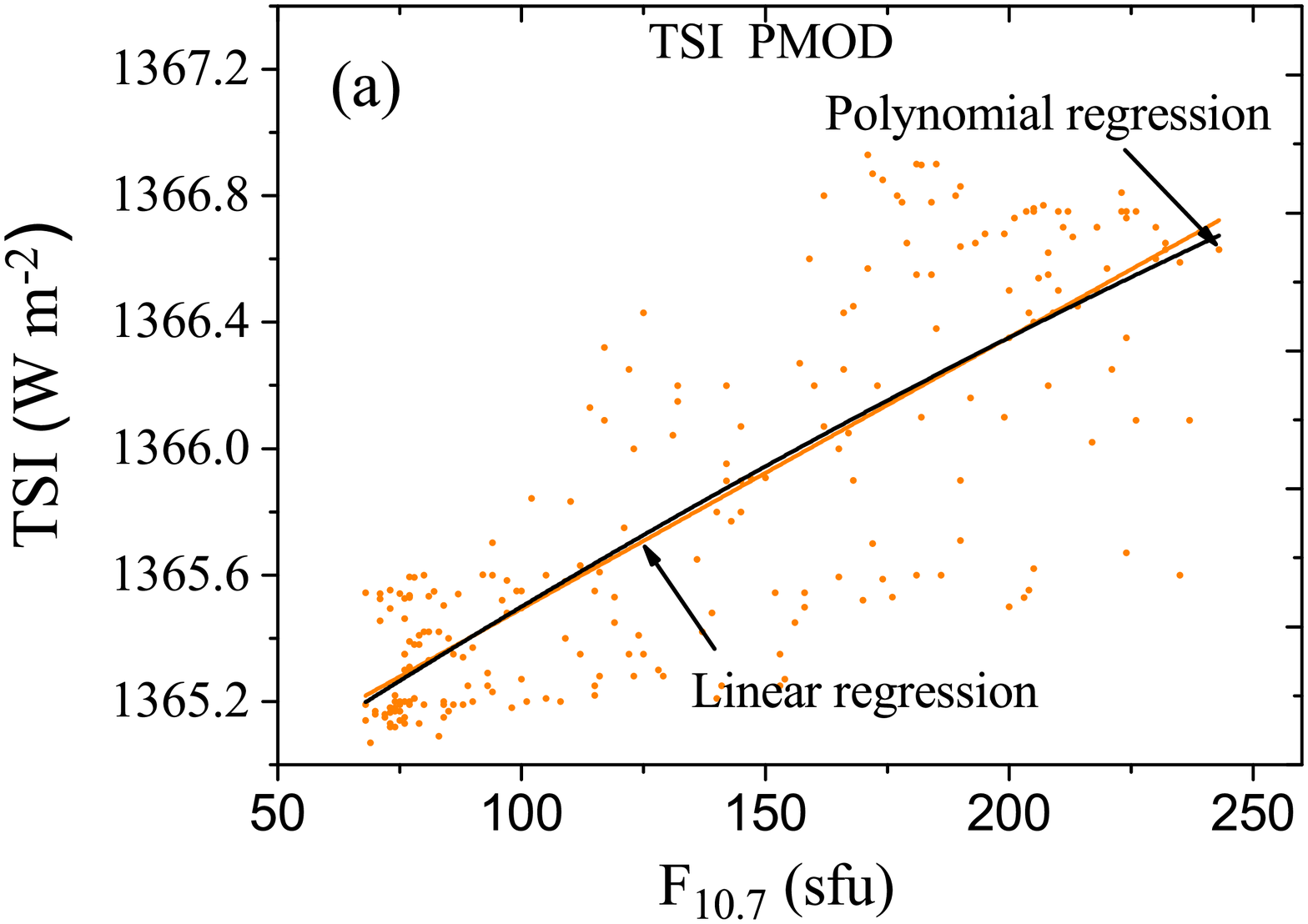}
               \hspace*{-0.015\textwidth}
               \includegraphics[width=0.555\textwidth,clip=]{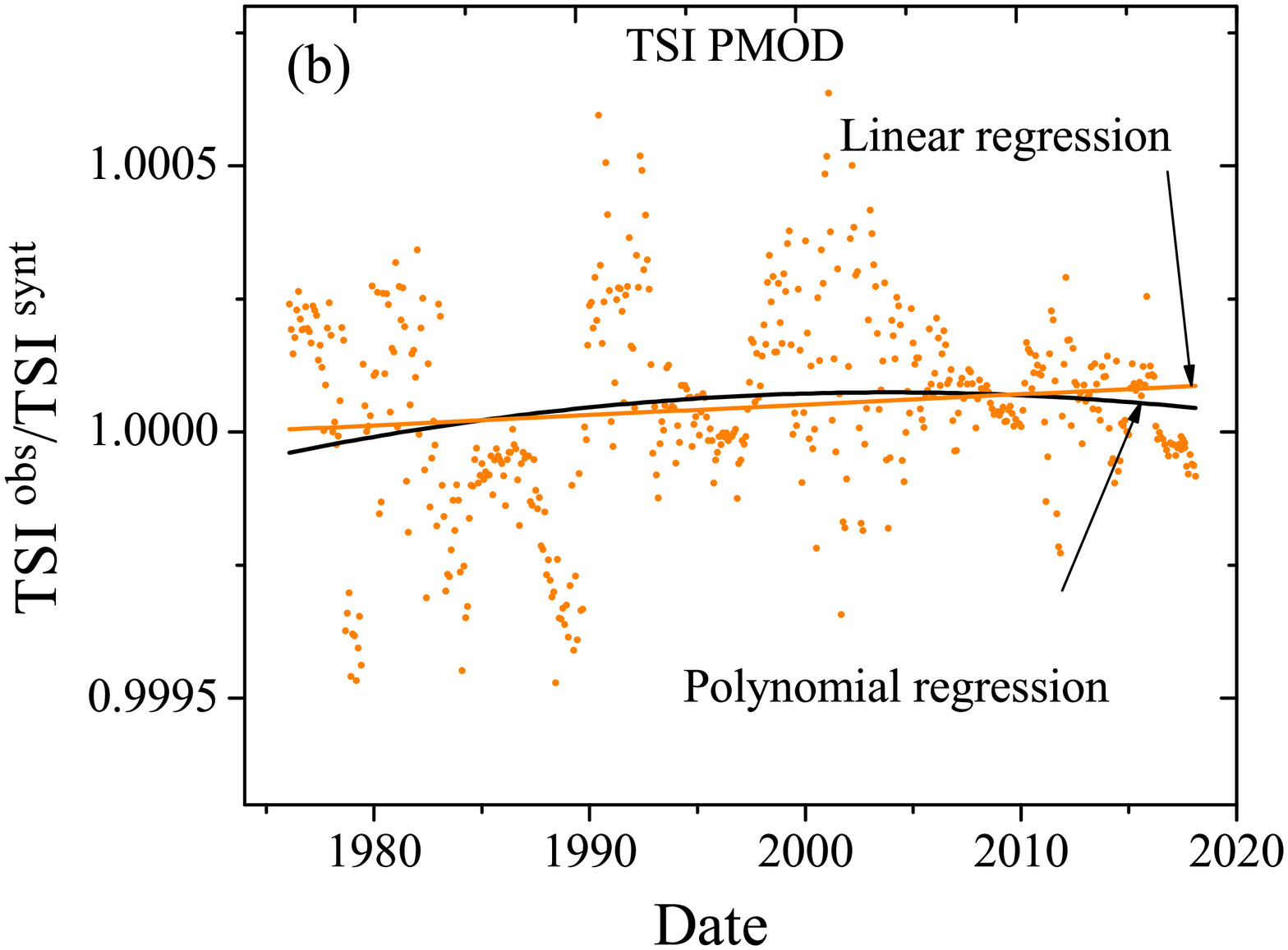}
              }
     \vspace{-0.35\textwidth}   
     \vspace{0.40\textwidth}    

        \caption{(a) $ TSI~PMOD^{obs}$ versus $F_{10.7}$ for the period from 1978 to 1990;
 (b) AIFF of TSI~PMOD composite irradiance (TSIPMFF)-- $TSI ~PMOD^{obs}/
 TSI~PMOD^{synt}$ for the period from 1978 to 2017.
        }
   \label{F-3panels}
\end{figure}

Measurements of TSI made with different equipment on different satellites have systematic errors related to special features of different measuring instruments and to their calibrations.

TSI measurements were carried out on the Nimbus-7 space satellite (1978 -- 1993)in the framework of the SMM mission on the satellite ACRIM-I (1980 -- 1989), on the ERBS satellite (1984 -- 1995), on the NOAA  satellites (1985-1989), on the ACRIM-II and the  ACRIM-III satellites (1991 -- 2003).

In recent time, instruments on space satellites of ACRIM series in addition to VIRGO and SoHO successfully continue observations of TSI. Observations on these instruments track variations in TSI with an amplitude of about 0.2\% when sunspots travel through a solar disk, while long variations in the solar cycle are only 0.1\%. SORCE TSI data sets achieved the improved accuracy of ±0.035\% (Dudok et al. 2017).

Inter comparisons of different data have led to some conclusions.
Willson et al. (1997) combined the SMM/ACRIM-I data with the later
UARS/ACRIM-II data by using inter comparisons with Nimbus-7 and ERBS
and concluded that the Sun was brighter by about 0.037 \% during the
minimum of cycle 22 than it was during the cycle's 21 minimum
(so-called "the trend between minima").

We used two composite data: 

(1) PMOD composite uses Nimbus7 and
ACRIM 1 -- 3 data on original VIRGO scale, datasets are available in Web-site, see Table 1.

(2) ACRIM composite also uses Nimbus7 /ERB, ACRIM 1 -- 3, datasets are
available in Web-site, see Table 1.

 Monthly plot of PMOD composite data is presented in Figure 2e.
 
In Figure 8a we can see that a relationship between the TSI and
the $F_{10.7}$ is not so close as between other solar Activity Indices and the $F_{10.7}$:
a correlation coefficient for monthly TSI versus the $F_{10.7}$ is equal to
0.76 while the correlation coefficients for other AI versus
the $F_{10.7}$ (in the same period 1950 -- 1990) are equal to 0.90 --
0.96.

Figure 8b shows that the $TSI~PMOD^{obs}/TSI~PMOD^{synt}$ (TSIPMFF) long-term trends.

We can see that TSIPMFF remained a relatively constant with
very weak positive trend (like for normalized $F_{Ly-\alpha}$) compared to previous cycles 22 -- 23. This trend differs from the
normalized SSN and normalized $F_{530.3}$ trends, in which the AIFF significantly decreased. 

The trend which is described in Figure 8b by
a linear regression shows an slight increase in the level of TSIPMFF
 ($TSI~PMOD^{obs}/TSI~PMOD^{synt}$) opposite the SSF ($SSN^{obs}/ SSN^{synt}$) trend when values 
reduces from 1 to 0.6. (Svalgaard 2013) suggested that the marked
reduction of the total area of spots in recent times has lead to an
increase of TSI. This small effect can be seen in Figure 8b.

\begin{figure}    
   \centerline{\hspace*{0.015\textwidth}
               \includegraphics[width=0.56\textwidth,clip=]{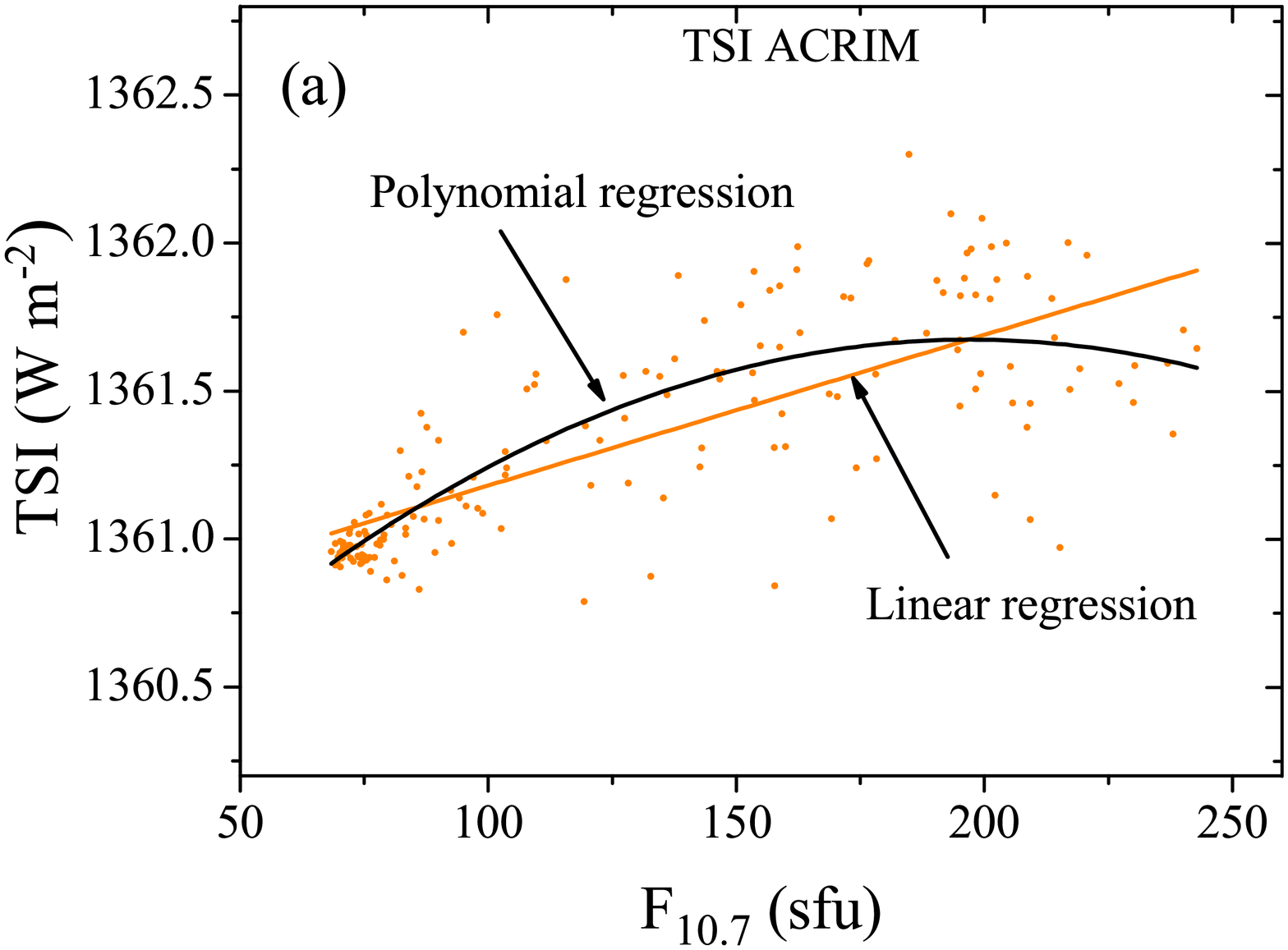}
               \hspace*{-0.1\textwidth}
               \includegraphics[width=0.58\textwidth,clip=]{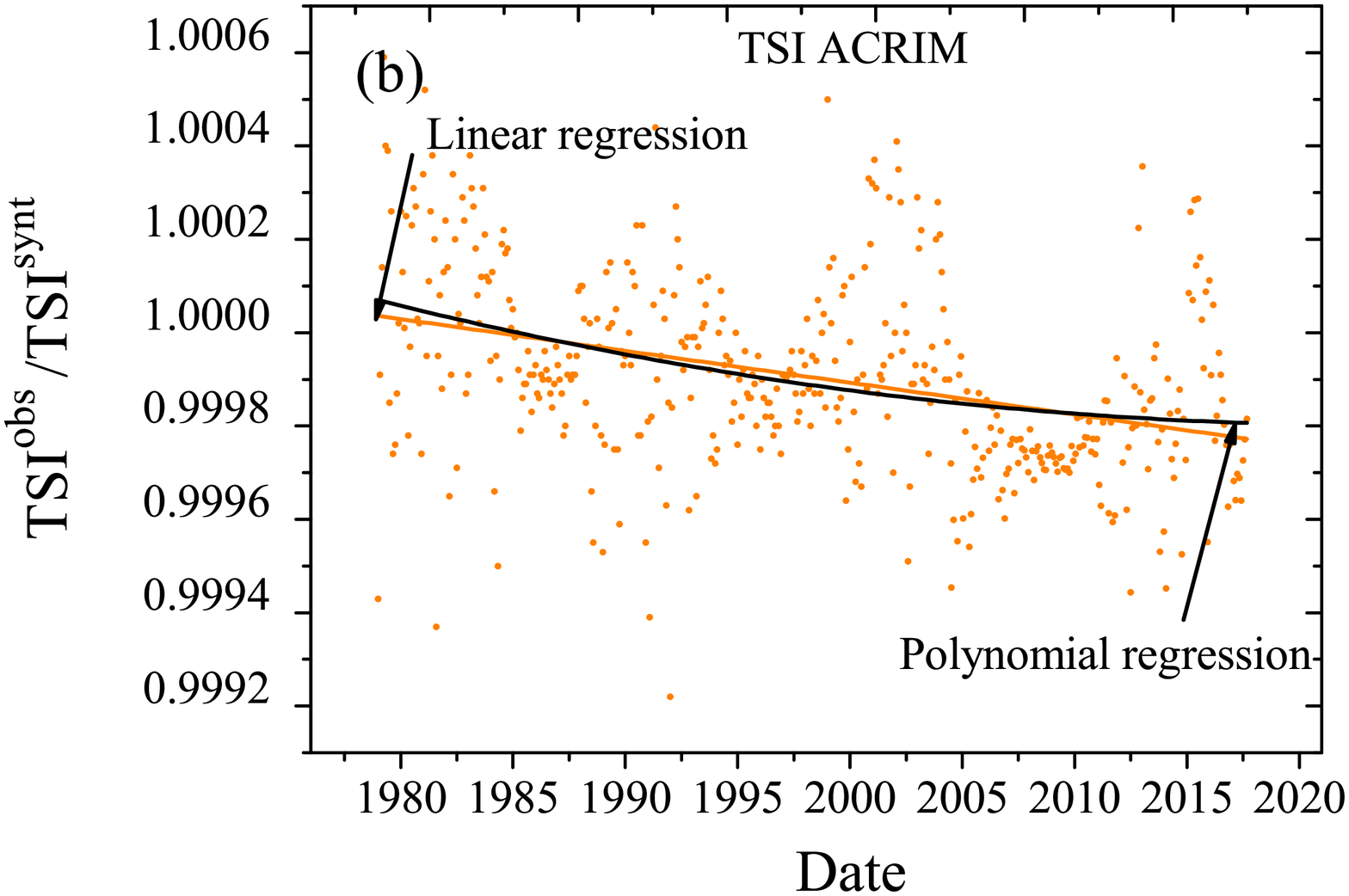}
              }
     \vspace{-0.35\textwidth}   
     \vspace{0.40\textwidth}    
   \caption{(a) $TSI~ACRIM^{obs}$ versus $F_{10.7}$ for the period from 1975 to 1990;
 (b) AIFF for the TSI ACRIM irradiance (TSIACRFF) -- $TSI~ACRIM^{obs}/TSI~ACRIM^{synt}$ for the period from 1975 to 2017.
        }
\label{F-2panels}
\end{figure}

In Figure 2f we present data of ACRIM composite based
on Nimbus7,
 ACRIM-1, ACRIM-2, VIRGO and ACRIM-3 observations.  
We use the TSI data set in cycle 24 obtained from these satellites which were collected in
NGDC and calibrated on SORCE/TIM level.

In Figure 9a we see that the relationship between the TSI and the $F_{10.7}$
is not very close and correlation coefficient of linear 
regression is equal to  0.73.

Figure 9b shows that the normalized index TSIACRFF -- 
$TSI~ACRIM^{obs}/TSI~ACRIM^{synt}$ became the normalizedly constant with a very weak
negative trend (as AIFF for  $F_{Ly-\alpha}$) unlike the normalized SSN and normalized 
$F_{530.3}$, for which the ratio AIFF became significantly lower in cycle 24. 

 We have examined data of the TSI in
two main versions, and can state that long-term trends in both cases
are slightly different.

\subsection{Flare Index}

The Flare Index is a measure of a short-lived activity on the Sun
and this index is approximately proportional to the total energy emitted by a
flare: $FI = i\cdot t $, where $i$ is the intensity scale of
importance and $t$ is the duration of the flare in minutes. 
Flare activity (according to the FI monthly variations) was very high in cycles 21 and 22. In cycle 23 the flare activity was twice lower. And in 24th cycle, the flare activity was half as much as in 23rd cycle. A study and forecasts of the FI are  necessary to evaluate the phenomenon of flare activity, since large flares lead to solar proton events, the largest of which lead to strong magnetic storms in the Earth's magnetosphere, to disruption of radio communication and to threaten the health of astronauts.

For the FI study we used NASA datasets which are available in Web-site, see Table 1.

\begin{figure}    

   \centerline{\hspace*{0.015\textwidth}
               \includegraphics[width=0.55\textwidth,clip=]{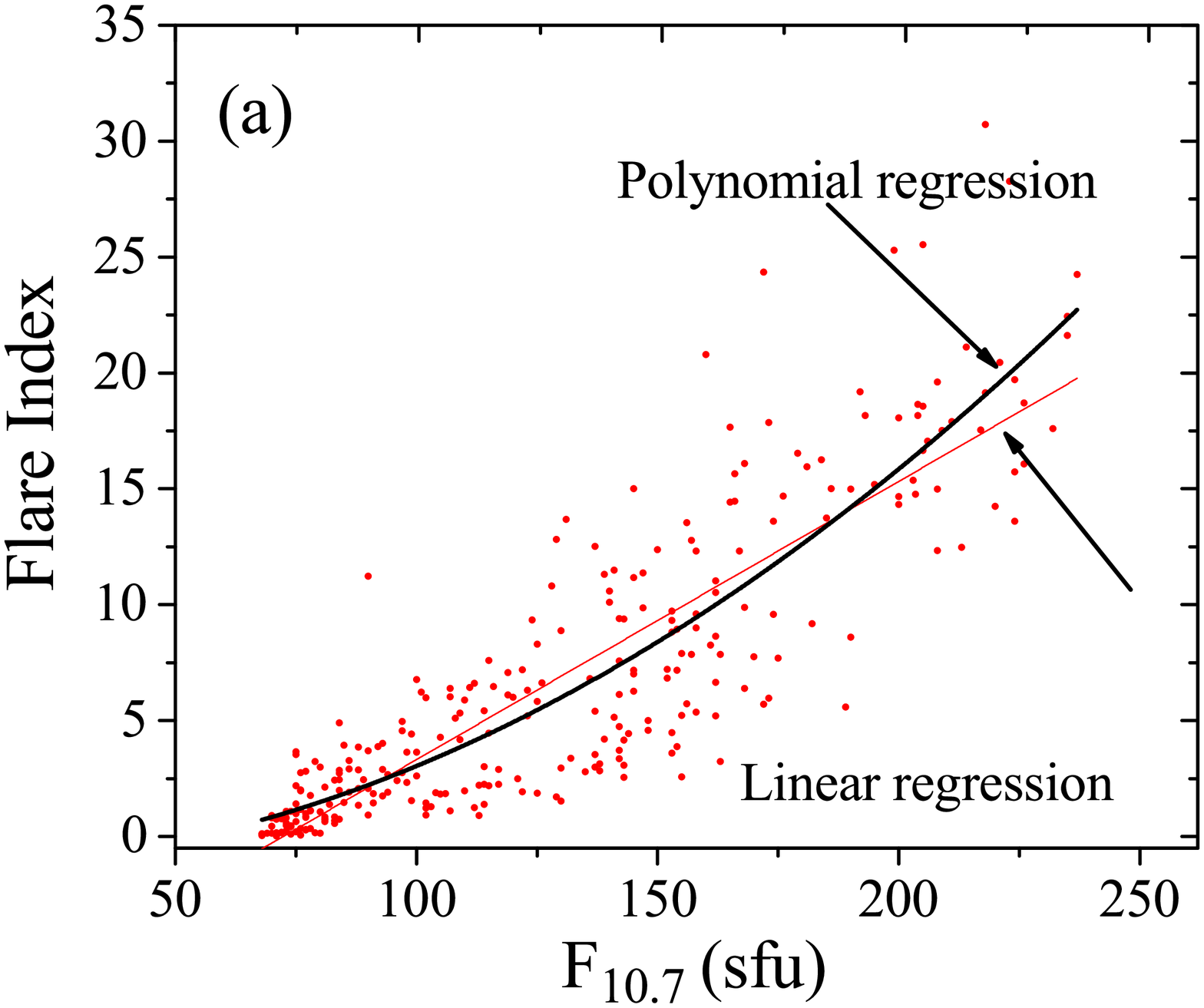}
               \hspace*{-0.015\textwidth}
               \includegraphics[width=0.6\textwidth,clip=]{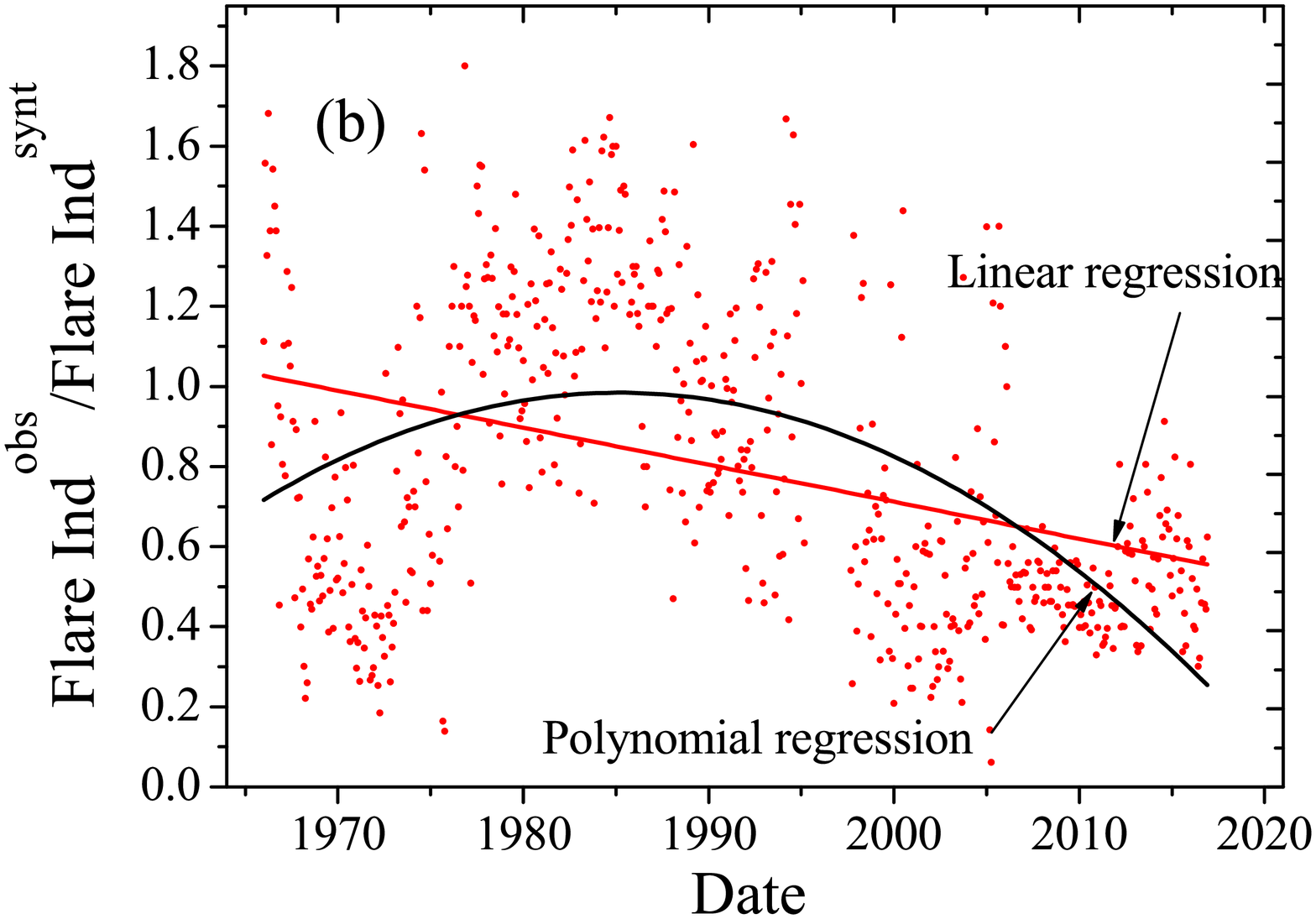}
              }
     \vspace{-0.35\textwidth}   
     \vspace{0.45\textwidth}    

\caption{(a) $FI^{obs}$ versus $F_{10.7}$ for the period from 1965 to 1990;
 (b) AIFF for the Flare Index (FIFF) -- $FI^{obs}/FI^{synt}$ for the period from 1965 to 2017.
        }

   \label{F-2panels}
\end{figure}

Atac \& Ozgus (1998) compared the FI in 
cycle 22 with similar solar activity indices (the SSN, the $F_{530}$, the
$F_{10.7}$, the TSI that arise under different physical conditions) for 
learning how the FI agrees with other solar indices. It was pointed that the FI is  well indicator of solar activity on short (minutes to hours) time-scales. Daily and monthly FI well correlates with daily and monthly SSN, $F_{10.7}$ and $F_{530}$. But connection between the FI and the TSI is not good enough.

In Figure 2g we show NASA data for the FI until 2014.
Note than for the period 2015 to 2017, we have calculated the FI using the information from the NASA catalog of flares which has information both of flares in optical range and of flares in X-ray range. This catalog of flare events is available in Web-site, see  
http://www.wdcb.ru/stp/data/Solar\_Flare\_Events/Fl\_XXIV.pdf.

In Figure 10a we show
monthly FI versus the $F_{10.7}$  during the period of stability 1950 -- 1990.
A correlation coefficient of a linear regression is equal to
0.87. It is high enough in comparison with the TSI versus the $F_{10.7}$ correlation.

A time dependence of the normalized FI (FIFF) in Figure 10b shows the considerable
long-term trends in FIFF.
According to the linear trend  we see that the FIFF has 
reduced from 1 in 1970 -- 1990 to 0.5 in 2017 but a polynomial
trend show the more significant reduce from 0.9 -- 1.1 in 1970 -- 1990
to 0.35 in 2017.

\section{Discussion} 
      \label{S-Discussion}
      
Our analisis shows that the determination and further our study of normalized AI -- AIFF has been very useful for the study and for the subsequent conclusions about the behaviour of various AI over the 21 -- 24 activity cycles.

Note that the normalized AIFF indices should be compared with the ideal case when these values are close to 1. The noticeable differences of AIFF from 1 indicate that there are constant trends in changing of the AI values as long-term trends (on  which the linear regression points out) and as short-term trends (on which the polynomial regression points out) takes place.

The study of long-term variations in the SSN-index showed that the current 24th cycle may be a precursor of a minimum of solar activity, similar to the minimum of Dalton.
 This minimum of solar activity was observed in the late 18th -- early 19th century during 3 cycles one after another, see Figure 1. Thus, if cycles 25 and 26 will be the same as cycle 24 or will be the less, the current minimum in solar activity may be similar to the minimum of Dalton.
For example in (Gopalswamy et al. 2018) predicted that cycle 25 will not be too different from cycle 24 in its strength.  This prediction for cycle 25 is based on a long-term solar activity studies using microwave imaging observations from Nobeyama Radiohelioograph at 17 GHz.

We can see that variations in AI which show the most decreases of normalized indices AIFF are connected to the temporary variations of large-scale magnetic fields in the photosphere (the SSN) and in the corona (the $F_{530.3}$ and the Flare Index). So, the greatest changes were link to those activity indices that are most closely related to local magnetic fields. 
The number of sunspots directly depends on the number of outputs of strong magnetic fields into the atmosphere of the Sun from the convective zone under the photosphere. So, If the activity of the local magnetic field is reduced, this leads to decreasing the index of the SSN (normalized index -- SSFF).
 The coronal index $F_{530.3}$ (normalized index -- 530FF) is also close connected to the magnetic fields because of all the substance in corona is concentrated along the magnetic field lines. The growth of the intensity of the magnetic field lines leads to an increase in the density and concentration of the corona substance, which leads to an increase in the 530.3 nm coronal flux and vice versa.
The Flare Index (normalized index -- FIFF)is also closely related to magnetic fields, since the flash most often comes from active areas (characterized by spots in the photosphere) in which the magnetic flux of the opposite sign emerges. As a result of the entanglement of the magnetic configuration of the spot, the probability of reconnection of magnetic lines in the tops of magnetic loops increases, which leads to flare release of energy.

The fact which is also important in our analysis is that for the evaluation of total level of radiation of the Sun the magnitude of the flux at 10.7 cm is chosen.
The values of $F_{10.7}$ are available on the Web-site (see Table 1) almost in real time. Measurements of this index are made with high accuracy and constant calibration, which is necessary for carrying out of a comparative analysis.

The $F_{10.7}$ index changes significantly (in 4 -- 5 times) during one cycle of activity (unlike, for example, the TSI which changes in 0.1 -- 0.15 \%), and this is convenient for calculations and improves the interpretation of the analysis done.

\section{Conclusion} 
      \label{S-Conclusion}

The analysis of different solar AI carried out in this work showed  the existence of short-term and long-term trends in AI time dependences over last 40 years.

1. The trends of recent time (1990 -- 2017) demonstrate for normalised values of the SSN, the FI and the $F_{530.3}$  the existence of sharp decrease of their normalised values SSFF, FIFF and 530FF. We see that together with reductions of their absolute values the examining their normalized variations AIFF (which are defined as $ F_{ind}^{obs}/F_{ind}^{synt}$ )
there exists an additional reduction of the SSN, the FI and the $F_{530.3}$. This additional  reduction have to be taken into account and to be analysed summarizing with the expected reduction which have been 
connected with cyclic (11-yr, half century and century) variations in solar activity. 

2. The most values of additional (compared to 1) decreases of normalized indices AIFF over last 40 years show the next AI: (1) SSN (SSFF),
is approximately decreased to 20\%, (2)  
$F_{530.3}$ (530FF) the
additional decrease is approximately equal to 30\%, (3) Flare Index (FIFF) the
additional decrease is approximately equal to 50 -- 70 \%
as for linear and for polynomial regression models.

3. Recent trends for TSI and UV-indices (MgII c/w and
the $F_{Ly-\alpha}^{obs}$) demonstrate that their normalised values ($F_{ind}^{obs}/F_{ind}^{synt}$ -- AIFF) show the approximately constant
behaviour during the period 1990 -- 2017.

4. For the
$TSI~PMOD$ composite we see a slight increase of TSIPMFF simultaneously with noticeable 
decrease of SSFF. This confirms an expectation of (Svalgaard 2013)
when with becoming fewer in size of areas of dark spots on solar disc, the TSI should be a little larger.

Acknowledgements goes to the to the science team of the Institute of Environmental Physics,
University of Bremen for MgII c/w composite data; to the ACRIM and PMOD science teams
for their corresponding TSI composites; to the Ottawa/Penticton team; to the LASP Composite Lyman Alpha science team; to the science team of Slovak Academy of Sciences for the green corona intensity.

\end{document}